\documentclass[usenatbib]{mn2e}
\usepackage{epsfig}
\newcommand{\hb}{H$\beta$}
\newcommand{\hg}{H$\gamma$}
\newcommand{\hst}{{\it HST}}
\newcommand{\kms}{km~s$^{-1}$}
\newcommand{\mgb}{Mg$_b$}

\def\ga{\mathrel{\mathchoice {\vcenter{\offinterlineskip\halign{\hfil
$\displaystyle##$\hfil\cr>\cr\sim\cr}}}
{\vcenter{\offinterlineskip\halign{\hfil$\textstyle##$\hfil\cr
>\cr\sim\cr}}}
{\vcenter{\offinterlineskip\halign{\hfil$\scriptstyle##$\hfil\cr
>\cr\sim\cr}}}
{\vcenter{\offinterlineskip\halign{\hfil$\scriptscriptstyle##$\hfil\cr
>\cr\sim\cr}}}}}
\def\la{\mathrel{\mathchoice {\vcenter{\offinterlineskip\halign{\hfil
$\displaystyle##$\hfil\cr<\cr\sim\cr}}}
{\vcenter{\offinterlineskip\halign{\hfil$\textstyle##$\hfil\cr
<\cr\sim\cr}}}
{\vcenter{\offinterlineskip\halign{\hfil$\scriptstyle##$\hfil\cr
<\cr\sim\cr}}}
{\vcenter{\offinterlineskip\halign{\hfil$\scriptscriptstyle##$\hfil\cr
<\cr\sim\cr}}}}}

\title[Evolutionary status of early--type galaxies in clusters at $z\approx0.2$ -- I.]{On
the evolutionary status of early--type galaxies in \\ clusters at
$z\approx0.2$ -- I. The Fundamental Plane\footnotemark[2]}

\author[A. Fritz et al.]
{Alexander Fritz$^{1}$\footnotemark[1],
 Bodo L.\ Ziegler$^{1}$,
 Richard G.\ Bower$^{2}$,
 Ian Smail$^{3}$, \and
 Roger L.\ Davies$^{4}$ \vspace{1mm}\\
$^{1}$Universit\"ats--Sternwarte G\"ottingen, Geismarlandstra{\ss}e 11,
D-37083 G\"ottingen, Germany \\
$^{2}$Department of Physics, University of Durham, South Road,
Durham DH1~3LE, UK \\
$^{3}$Institute for Computational Cosmology, University of Durham,
South Road, Durham DH1~3LE, UK \\
$^{4}$University of Oxford, Astrophysics, Denys Wilkinson Building,
Keble Road, Oxford OX1~3RH, UK}

\date{Accepted. Received --; in
  original form 2004 }
   
\pagerange{\pageref{firstpage}--\pageref{lastpage}} \pubyear{2004}

\begin{document}

\maketitle

\label{firstpage}

\begin{abstract}
We investigate a spectroscopic sample of 48 early-type galaxies in the rich
cluster Abell~2390 at $z=0.23$ and 48 early-type galaxies from a previously
published survey of Abell~2218 at $z=0.18$. The spectroscopic data of A\,2390
are based on Multi-Object-Spectroscopy with MOSCA at the 3.5-m telescope on
Calar Alto Observatory and are complemented by ground-based imaging
with the 5.1-m Hale telescope and \textit{Hubble Space Telescope} (\hst)
observations in the F555W and F814W filters. Our investigation spans a broad
range in luminosity ($-20.5\geq M_{r}\geq-23.0$) and a rather wide
field-of-view of $1.53\ h^{-1}_{70}\times 1.53\ h^{-1}_{70}$~Mpc$^{2}$.

Since the A\,2218 and A\,2390 samples are very similar, we can combine them
and analyse a total number of 96 early-type (E+S0) galaxies at $z\sim 0.2$.
Using the ground-based data only, we construct the Faber--Jackson relation (FJR)
for all 96 E+S0 galaxies and detect a modest luminosity evolution with
respect to the local reference. The average offset from the local FJR in the
Gunn $r$-band is $\Delta\,\overline{M}_{r}=0.32\pm0.22^{m}$. Similar results
are derived for each cluster separately.
Less-massive galaxies show a trend for a larger evolution than more-massive
galaxies. HST/WFPC2 surface brightness profile fits were used to derive the
structural parameters for a sub-sample of 34 E+S0 galaxies. We explore the
evolution of the Fundamental Plane (FP) in Gunn~$r$, its projections onto the
Kormendy relation and the $M/L$ ratios as a function of velocity dispersion.
The FP for the cluster galaxies is offset from the local Coma cluster FP. At a
fixed effective radius and velocity dispersion our galaxies are brighter than
their local counterparts. For the total sample of 34 E+S0 cluster galaxies
which enter the FP we deduce only a mild evolution with a zero-point offset of
0.10$\pm$0.06, corresponding to a brightening of 0.31$\pm$0.18$^{m}$.
Elliptical and lenticular galaxies are uniformly distributed
along the FP with a similar scatter of 0.1~dex. 
Within our sample we find little evidence for differences between the
populations of elliptical and S0 galaxies. There is a slight trend that
lenticulars induce on average a larger evolution of
0.44$\pm$0.18$^{m}$ than ellipticals with 0.02$\pm$0.21$^{m}$.
The $M/L$ ratios of our distant cluster galaxies at $z=0.2$ are offset
by $\Delta\,{\rm log}\,(M/L_{r})=-0.12\pm 0.06$~dex compared to those of
Coma. Our results can be reconciled with a passive evolution of
the stellar populations and a high formation redshift for the bulk of the
stars in early-type galaxies. However, our findings are also consistent with
the hierarchical formation picture for rich clusters, if ellipticals in
clusters had their last major merger at high redshift.
\end{abstract}

\begin{keywords}
galaxies: elliptical and lenticular, cD -- galaxies: evolution
-- galaxies: fundamental parameters -- galaxies: kinematics and dynamics 
-- galaxies: clusters: individual: Abell 2218
-- galaxies: clusters: individual: Abell 2390
\end{keywords}

\footnotetext[1]{E-mail: afritz@uni-sw.gwdg.de}

\footnotetext[2]{Based on observations collected at the Centro
  Astron\'omico Hispano Alem\'an (CAHA) at Calar Alto, operated by the
  Max-Planck-Institut f\"ur Astronomie, Heidelberg, jointly with the
  Spanish National Commission for Astronomy.}


\section{Introduction}\label{intro}

Since the first development and application of theoretical galaxy models
(e.g., \citeauthor{Lar:75} \citeyear{Lar:75};
\citeauthor{Too:77} \citeyear{Too:77}) much progress has been achieved to
explore the evolution of early--type (E+S0) galaxies. Nevertheless, one of the
key questions of early-type galaxy formation and evolution has not been fully
answered yet: When and within what time-scales have the stellar populations of
early--type galaxies been formed?

In the hierarchical galaxy formation picture the formation histories of
early-type galaxies in cluster and low-density environments are significantly
different (\citeauthor{BCF96} \citeyear{BCF96};
\citeauthor{KS:98} \citeyear{KS:98}). Merging of galaxies and the infall of new
cold gas from the galaxy haloes are the main drivers of structure evolution.
Clusters are formed out of the highest peaks of primordial density fluctuations
and most stars of massive cluster ellipticals are generated at high redshift
($z\ge 2$), which is in agreement with observational findings (e.g.,
\citeauthor{ZB97} \citeyear{ZB97}; \citeauthor{DFKI98} \citeyear{DFKI98}).
However, explicit age variations between early-type cluster galaxies and E+S0
galaxies in low-density regions are predicted. For clusters, models predict
ellipticals to have a mean luminosity-weighted age of 9.6~Gyr and lenticular
galaxies to be younger by $\sim$1~Gyr (\citeauthor{BCF96} \citeyear{BCF96};
\citeauthor{Col:00} \citeyear{Col:00}).
Both types indicate a weak trend that fainter galaxies
are older. On the contrary, for early-type galaxies in low-density regions
the hierarchical cluster models predict a broader age spread over a larger
luminosity range and mean luminosity-weighted ages of $\sim$5.5~Gyr.

Numerous observational studies imply four main evidences for a high redshift
formation of early-type galaxies: ({\it i}\/) the homogenous stellar populations
of E+S0 galaxies seen in tight correlations between colours/absorption
line strengths and velocity dispersion and in the Fundamental Plane (FP),
which combines kinematics with structural properties, ({\it ii}\/) the 
[Mg/Fe] overabundance and ({\it iii}\/) the weak evolution in colours and
line strengths with redshift.

In the nearby Universe, inconsistent results have been acquired regarding
any possible difference between field and cluster galaxies. For example,
\cite{deCD:92} derived from a subset of cluster and field early-type galaxies
taken from the ``Seven Samurai'' group (\citeauthor{Fab:89} \citeyear{Fab:89};
\citeauthor{DD87} \citeyear{DD87}) that field ellipticals show a
larger scatter in their properties indicating that they consist of younger
stellar populations than cluster galaxies.
\cite{JM99} investigated NIR spectra of 50 ellipticals in three nearby
clusters and in the field, using the CO (2.3$\mu$m) absorption
feature to explore the presence of an intermediate-age population.
They detected no stronger CO absorption for the field ellipticals.
Very isolated field ellipticals show a very homogenous population and
a small range of metallicity with no sign of recent star formation (SF).
In groups, ellipticals have a wide range in metallicity, mostly showing
evidence for an intermediate-age population,
whereas in rich clusters they exhibit intermediate properties in 
metallicity and CO absorption.
\cite{Ber98} analysed a large sample of ENEAR field and cluster galaxies and
found slight zero-point changes in the Mg$_{2}-\sigma$ relation. They
explain this as an age difference, with field objects being younger by
$\sim$1~Gyr. However, they conclude that the bulk of stellar populations of
E+S0 in both environments has been formed at high redshifts ($z\ga 3$).
\cite{KSCD02} detected in a sample of nine local early-type galaxies (five
morphologically disturbed) in low-density environments (LDR) no strong ongoing
SF. The results were compared to cluster E+S0s in Fornax. The ages
of the LDR galaxies are spread over a broad distribution, similar to that of
Fornax S0 galaxies and being on average younger by 2-3 Gyr than the E+S0s 
in Fornax. These LDR galaxies indicate 0.2~dex \textit{higher} metallicities
(in conflict with semi-analytical models) than their cluster representatives,
which suggests that the formation of E+S0 galaxies in low-densities continues
to $z\la 1$, whereas in clusters most stars have already been generated at
$z\ga 2$.
Recently, \cite{SBGCCG03} studied 98 E+S0 galaxies in the field and in clusters
and found higher C4668 and CN$_{2}$ absorption line strengths for the field
population. They interpret this as a difference in abundance ratios arising
from different star formation histories. However, both field and cluster
E+S0s show similar relations in \mgb$-\sigma$ and $\langle Fe\rangle-\sigma$.

At higher redshift differences between field and cluster galaxies should
become more apparent.
Recent results from investigations based on the Fundamental Plane at
intermediate redshift ($z\leq 0.5$), indicate no significant variations between
the cluster and field early-type populations
(\citeauthor{vD01} \citeyear{vD01}; \citeauthor{Treu01b} \citeyear{Treu01b};
\citeauthor{RKFKM03} \citeyear{RKFKM03}).
With respect to the mean age of these populations, field galaxies seem to comprise
slightly younger stars than the cluster population, whereas the majority of
stars must have formed at a much higher redshift of $z_{{\rm f}}>2$.
However, at higher redshift ($z\sim 0.7$), some studies derive a
significant offset between field and cluster galaxies 
(\citeauthor{TSCMB02} \citeyear{TSCMB02}).

Since the last years a multiplicity of investigations of distant rich clusters
have been performed
(\citeauthor{E97} \citeyear{E97}; \citeauthor{DOCSE97} \citeyear{DOCSE97};
\citeauthor{Sta:98} \citeyear{Sta:98}; \citeauthor{DFKI98} \citeyear{DFKI98};
\citeauthor{KIDF00} \citeyear{KIDF00}; \citeauthor{vD00} \citeyear{vD00};
\citeauthor{Z2001} \citeyear{Z2001}; \citeauthor{Treu03} \citeyear{Treu03};
\citeauthor{WvDKFI04} \citeyear{WvDKFI04}).
Most of these studies can be reconciled
with the picture of a monolithic collapse with a high redshift formation of
the stellar populations of E+S0 galaxies.
Results from these distant clusters have not found any differences in
the properties of E+S0 galaxies (e.g., \citeauthor{KIDF00} \citeyear{KIDF00}).
Recently, in a re-analysis of two high redshift clusters at $z=0.58$ and
$z=0.83$ no environmental dependence of the FP residuals was detected
(\citeauthor{WvDKFI04} \citeyear{WvDKFI04}).
When looking at the residuals of the FP, and suggesting that the
residuals correlate with environment, it is difficult to distinguish if this
effect is due to changes in velocity dispersion, size or luminosity of the
galaxies. Selection effects have strong influence on the parameters and can
also mimic possible correlations.
In a study on $\sim$9000 early-type galaxies from the SDSS
(\citeauthor{SDSSIIIFP03} \citeyear{SDSSIIIFP03}), a weak correlation between
the local density and the residuals from the FP was revealed, in the sense that
the residuals in the direction of the effective radii increase slightly as
local density increases. However, the offset is quite small and subject to
selection and evolutionary effects. The open question still to address is,
how this dependence occurs.

Looking at the morphology, the formation and evolution of lenticular galaxies
is different and stands in contrast to elliptical galaxies. Deep studies of
galaxies in distant rich clusters using the \textit{WFPC2} camera onboard
the \textit{Hubble Space Telescope (HST)} revealed that S0 galaxies show a
rigorous evolution with redshift in these dense environments
(e.g., \citeauthor{DOCSE97} \citeyear{DOCSE97}). Although S0 galaxies form the
dominant population in local rich clusters of $\sim$60\%, at intermediate
redshift ($z\sim0.5$) spiral and disturbed galaxies compose the major part of
the luminous galaxies, whereas S0 galaxies are less abundant (10--20\%).
\cite{SLCE99} studied early-type field
galaxies at intermediate redshifts ($z\sim0.5$) and detected
$[{\rm OII}]\lambda3727$ emission lines in about $1/3$ of these
galaxies, which indicates ongoing star formation. Furthermore, in about the
same fraction of faint spheroidal HDF galaxies significant variations of
internal colours were found, frequently showing objects with blue cores
(\citeauthor{MAE:01} \citeyear{MAE:01}).
The authors conclude that at $z\sim 1$ about half of the field S0 galaxies
show clear signs of star formation activity.
Using deep optical and NIR imaging, \cite{SKKSPFH01} revealed 
differences between more luminous ($\ga 0.5L^{*}_{K}$)
and less luminous ($\la 0.1L^{*}_{K}$) early-type galaxies in A\,2218.
The faintest S0s show a wide spread in the colours, and $\sim$30\,\% of these
S0s exhibit on average younger ages (2-5 Gyrs).

These results seem to imply that galaxy transformation via interaction 
is an important phenomenon in clusters. Due to the large
velocity dispersion mergers are less frequent in rich clusters, whereas
effects such as ram--pressure stripping by the hot intra cluster medium (ICM)
or tidal interactions between the galaxies are more likely.
A unique mechanism for the transformation into S0 galaxies is still
missing to explain the strong decrease in the frequency of S0's since the last
5\,Gyrs ($z\sim 0.5$). A possible scenario
is that field spiral galaxies falling into the cluster centre
experience a starburst phase, resulting in the Butcher--Oemler effect.
Ram--pressure stripping by the ICM (also maybe through tidal stripping)
over a short time-scale of less than one Gyr, could cause the wide-spread and
rapid decline in star formation leading to post--starburst galaxies and red
passive spiral galaxies (e.g., \citeauthor{BH92} \citeyear{BH92}). Harassment
by the tidal field of the galaxy cluster and high speed encounters have a non
negligible effect on the following passive evolution of a galaxy by removing
stars from the disk which may end up in an S0 galaxy (\citeauthor{MLKDO96}
\citeyear{MLKDO96}; \citeauthor{PSDCB99} \citeyear{PSDCB99}).

In terms of structural parameters, elliptical galaxies comprise not
a single homogenous group of galaxies but encompass two different groups,
\textit{disky} and \textit{boxy} ellipticals (\citeauthor{BDM88} \citeyear{BDM88};
\citeauthor{Ben88} \citeyear{Ben88}; \citeauthor{KB1996} \citeyear{KB1996}).
The shape of these galaxies is very important since it correlates with other
physical properties, such as luminosity, shape, 
rotation (axis) and core profile.
Recently, the origin of disky and boxy ellipticals was investigated
(\citeauthor{NB99} \citeyear{NB99}; \citeauthor{NB03} \citeyear{NB03}).
Equal-mass mergers result in an anisotropic system with slow
major axis rotation and a large amount of minor-axis rotation (boxy elliptical),
whereas unequal-mass merger of mass ratio 3~:~1 and 4~:~1 lead to a
rotationally supported system with only a small rotation along the minor-axis
(disky elliptical).
In general, giant \textit{high-luminous} ellipticals preferably contain
\textit{boxy} isophotes, whereas \textit{low-luminous} ellipticals comprise a
\textit{disky} structure. Could they maybe have a different evolution? At
intermediate redshift we cannot distinguish between disky and boxy galaxies.
However, with respect to our large sample we are able separate low from high
luminous galaxies and look for possible differences in their evolution. Results
of such a comparison would give conclusions if the two types of ellipticals
might undergo different formation scenarios.

Cluster environments provide the opportunity to observe a larger number of
early-type galaxies with multi-object spectroscopy simultaneously. Furthermore,
at a redshift of $z\sim 0.2$, the cluster galaxies are bright
enough to observe even sub-$L^{*}$ systems with 4-m class telescopes, while
still representing a look-back time of $\sim$3\,Gyrs, adequate to address
evolutionary questions. To look for the environmental dependence,
it is desirable to investigate possible radial dependences in age and
metallicity of stellar populations with a large sample of early-type galaxies.
Therefore, we have undertaken a programme to acquire high-quality spectra of
a large number of early-type galaxies in two rich clusters, $N=48$ for
Abell\,2218 and $N=48$ in the case of Abell\,2390, across a wide range in
luminosity (down to $M_{B}=M^{\ast}+1$, with $M^{\ast}=-19.5+5\,{\rm log}\,h_{70}$;
corresponding to $-19.2\geq M_{B}\geq-24.2$ for $z=0.23$) and a
wide field-of-view ($\sim$10$'\times 10'$). Each cluster centre has been
observed with \hst\ allowing therefore accurate structural parameter
determinations. As already demonstrated in \cite{Z2001}, both clusters may well
serve in the future as suitable benchmarks for the comparison to rich, high
redshift clusters, since aperture corrections are less crucial than e.g.
to the Coma cluster.

The analysis and results of our study of the cluster A\,2218 are discussed in
detail in \cite{Z2001} (hereafter Z01). In this article, we present
the investigation of A\,2390. Combining these two large data samples, a total
number of 34 E+S0 galaxies enter the FP, which represents -- apart from the
study of \cite{KIDF00} -- the most extensive investigation on early-type cluster
galaxies at intermediate redshift. Our study is dedicated to explore
intrinsically low-luminosity elliptical and S0 galaxies at redshifts where
some evolution is visible and over a wide field-of-view in order to look
for environmental variations of the evolution of E+S0 galaxies. Another
important aim of this work is also to break the ``age-metallicity'' degeneracy
of early-type galaxies. In a forthcoming paper we will explore the evolution of
the A\,2390 stellar populations in age, metallicity and abundance ratios by
analysing absorption line strengths (e.g., H$\beta$, Mg$_{b}$, Fe-indices), and
comparing them with stellar population models.

The paper is organised as follows. The photometry on the ground-based images
of A\,2390 and the {\it HST} structural analysis are presented in \S2.
Spectroscopic observations of the cluster A\,2390, the sample selection and
reduction of the spectra are described in \S3. In \S4 the evolutionary status
of the rich clusters A\,2390 and A\,2218 at $z\sim 0.2$ is illustrated. Results
for both clusters derived with scaling relations such as the Faber-Jackson
relation and FP and the evolution of the $M/L$ ratio are given in \S5.
A summary of our results is presented in~\S6.

In this paper we assume the concordance cosmology for a flat Universe with
$\Omega_{m}=0.3$, $\Omega_{\Lambda}=0.7$ and $H_0=70$\,km\,s$^{-1}$\,Mpc$^{-1}$.
This results for the nearby Coma comparison cluster ($z=0.024$) in a distance
modulus of $dm=35.10$\,mag, for the cluster A\,2218 ($z=0.175$) in
$dm=39.64$\,mag and for A\,2390 ($z=0.228$) in $dm=40.28$\,mag, a scale of
3.65\,kpc\,arcsec$^{-1}$ and a look--back time of $\sim2.75$\,Gyrs.


\section{Photometry of Abell~2390}

\subsection{Abell~2390}

The cluster Abell~2390 ($\alpha_{{\rm 2000}}=21^{{\rm h}} 53^{{\rm m}} 34\fs 6$,
$\delta_{{\rm 2000}}=+17^{\circ} 40\arcmin 10\farcs9$) at $z=0.228$, richness
class~1, has a large velocity dispersion,
$\sigma=1100\pm 63$\,\kms \citep{Car:96} and a high X-ray luminosity,
$L_{\rm X}$(0.7--3.5\,keV) = 4.7$\times 10^{44}$ erg\,s$^{-1}$ \citep{LeB:91}.
\cite{Car:96} analysed the dynamical state of the cluster and its mass
distribution and found a virial radius of $R_{v}=3.156\,h^{-1}_{100}$~Mpc and virial
mass of $M_{v}=2.6\times 10^{15} h^{-1} M_\odot$, which makes A\,2390 more
massive than Coma ($M_{v}=2.1\times 10^{15} h^{-1} M_\odot$). At a constant
mean interior density of 200$\rho_c$, A\,2390 and Coma have
$M_{200}$-masses of 1.2 and $1.3\times 10^{15} h^{-1} M_\odot$, respectively.

Our study of the early-type galaxy population in A\,2390 is based upon
Multi-Object Spectroscopy (MOS) using MOSCA ({\bf M}ulti {\bf O}bject
{\bf S}pectrograph for {\bf C}alar {\bf A}lto) at the Calar Alto 3.5-m
Telescope on Calar Alto Observatory (CAHA) in Spain (see~\S\ref{spec}).
In addition, optical photometry from the 5.1-m Hale telescope at Palomar
Observatory is available and we have exploited WFPC2 images taken with
\hst\ providing high-quality morphological information for a subset of our
sample. Table~\ref{tab-obs} gives a summary of the obtained observations.

{\scriptsize
\begin{table}
\begin{center}
\caption{\newline \centerline{\sc Log of Observations}} \label{tab-obs}
\begin{minipage}{\textwidth}
\vspace{0.1cm}
\begin{tabular}{lllcc} 
\hline\hline
\noalign{\smallskip}
Tel.      & Instrument &   Date  &  Band (phot)     &$T_{\rm exp}$\cr
          &            &         &  Mask / $N_{\rm spec}$ & [ksec] \cr 
\noalign{\smallskip}
\hline
\noalign{\smallskip}
HALE        & COSMIC  & 09--12/06/94 & $U$	&  3.00  \cr
            & COSMIC  & 09--12/06/94 & $B$	&  0.50  \cr
\smallskip  & COSMIC  & 09--12/06/94 & $I$      &  0.50  \cr
HST         & WFPC    & 10/12/94     & F555W	&  8.40 \cr
\smallskip  & WFPC    & 10/12/94     & F814W	& 10.50 \cr
CA\,3.5     & MOSCA   & 07--10/09/99 & 1 / 17	& 29.88 \cr
\smallskip  & MOSCA   & 07--10/09/99 & 2 / 22	& 42.12 \cr
            & MOSCA   & 26--28/07/00 & 3 / 24	& 42.48 \cr
\noalign{\smallskip}
\noalign{\hrule}
\end{tabular}
\end{minipage}
\end{center}
\end{table}
}

\subsection{Ground-based $UBI$ imaging}\label{gbima}

Abell~2390 was observed at the 5.1-m Hale telescope on Mount Palomar
using COSMIC ({\bf C}arnegie {\bf O}bservatories {\bf S}pectroscopic
{\bf M}ultislit and {\bf I}maging {\bf C}amera) in the $U$ (3000~sec),
$B$ (500~sec) and $I$-band (500~sec), allowing to select early-type
galaxies in the full field-of-view of MOSCA due to the nearly equally large
field-of-view of COSMIC of $9.7'\times 9.7'$.
Seeing conditions ranged from 1.4\arcsec\ in the $U$-band,
1.3\arcsec\ in the $B$ to 1.1\arcsec\ in the $I$-band
(\citeauthor{SEEB98} \citeyear{SEEB98}). At $I=22.5$~mag a completeness
level of 80\% from a comparison with deeper field counts is warranted.
All frames from the ground-based imaging data were reduced in a standard
manner with {\sc iraf}\footnote{IRAF is distributed by the National Optical
Astronomy Observatories, which are operated by the Association of
Universities for Research in Astronomy, Inc., under cooperative
agreement with the National Science Foundation.} using standard reduction
packages.

As a consistency check of our ground-based photometry we compared
our photometric data with the results for the cluster A\,2390 by \cite{CNOCII}
which were derived as part of the CNOC cluster redshift survey.
Through a cross-correlation we identified 12 galaxies with spectra which are
included in both data sets. After the transformation of our $I$ to Gunn
$r$ magnitudes, we found no significant difference between the magnitudes,
$\Delta (r-r_{{\rm Yee}})=0.04\pm0.16$~mag. In addition, the redshift
determinations of all these objects show a very good agreement.

Absolute magnitudes were calculated from our ground-based $UBI$ imaging.
Total apparent magnitudes were derived with the Source Extractor package
(SExtractor, see \citeauthor{BA96} \citeyear{BA96}) as given by {\tt Mag\_BEST}.
The transformation of the Johnson-Kron-Cousins $I$-band magnitudes to absolute Gunn $r$
rest-frame magnitudes was performed in the following way. Synthetic photometry
was performed using the observed spectral template for a typical E/S0 galaxy by
\cite{KCBMSS96} and the synthetic spectral templates by \cite{Moe01}, which
were generated with evolutionary synthesis models. The SEDs were redshifted to
the cluster redshift of $z=0.23$ to determine the flux through the $I$
filter and at $z=0$ through the \cite{TG76} $r$ filter. This lead to the
transformation $I_{{\rm obs}}-{\rm Gunn}\ r_{{\rm rest}}=-0.81$. 
Typical uncertainties in the k-corrections are $\Delta k_{r}=0.03^{\rm m}$
for the SEDs of both ellipticals and S0 galaxies.

The correction for the Galactic extinction was performed by using the
\textit{COBE} dust maps by \cite{SFD98}. For A\,2390 an index
$E(B-V)=0.110^{\rm m}$ was determined, resulting in extinction coefficients for
the Johnson Cousins filters of A$_{U}=0.600^{\rm m}$, A$_{B}=0.476^{\rm m}$
and A$_{I}=0.214^{\rm m}$, respectively.
For the \textit{HST/WFPC2} F814W filter $A_{814}=0.214^{\rm m}$ was derived. 
The uncertainty in the extinction is ${\rm E(B-V)}=0.010^{\rm m}$.

The total errors in absolute magnitude Gunn $r$, $\delta \sigma_{M_{r}}$,
follow as the linear sum of the errors in the total $I$-band magnitude
$\delta \sigma_{I}$, the uncertainty in the k-correction
$\delta \sigma_{k_{r}}$ and the error in the extinction correction
$\delta \sigma_{\rm A_{I}}$. For our ground-based magnitudes
$\delta \sigma_{M_{r}}$ covers the range between
$0.07^{\rm m}\leq\delta\sigma_{M_{r}}\leq0.14^{\rm m}$, with
an average error of $\delta\overline{\sigma}_{M_{r}}=0.09^{\rm m}$.


\subsection{Structural parameters}
\label{strucpara}

During Cycle~4 A\,2390 was observed with
the \emph{HST}\footnote{Based on observations made with the
NASA/ESA \emph{Hubble Space Telescope}, obtained at the Space Telescope Science
Institute, which is operated by the Association of Universities for Research in
Astronomy, Inc., under NASA contract NAS 5--26555. These observations are
associated with program \#\,5352.} in the filter F555W ($V_{555}$)
with 8400~sec and in the F814W ($I_{814}$) with 10500~sec as part of a large 
gravitational lensing survey. These exposure times are
deep enough to determine structural parameters down to
$B_{{\rm rest}}$$\sim$23 mag (\citeauthor{ZSBBGS99} \citeyear{ZSBBGS99}).

\subsubsection{Choice of luminosity profile}

A fundamental question when determining the structural parameters of galaxies
is what kind of surface brightness profile type is the most suitable. Generally,
derivation of total magnitudes involves an extrapolation of curves of growth to
infinity, thus relying on fits to the luminosity profiles of the galaxies.
For example, \cite{Cao93} used S\'{e}rsic profiles to fit the surface
brightness profile of early-type galaxies
(see also \citeauthor{LBBMMC02} \citeyear{LBBMMC02}). Another approach is to
use the classical de Vaucouleurs ($r^{1/4}$) law with an exponential disk
component (e.g., \citeauthor{SWVSP02} \citeyear{SWVSP02};
\citeauthor{TSIF03} \citeyear{TSIF03}).

\cite{TSIF03} explored the results of three different profile
models (double exponential, S\'{e}rsic bulge+exponential disk and
an $r^{1/4}$ bulge+exponential disk) using the GIM2D package
(Galaxy Image Two-Dimensional, \citeauthor{SWVSP02} \citeyear{SWVSP02})
on a sample of 155 cluster galaxies in CL\,1358+62 at $z=0.33$.
Based on their extensive profile tests these authors conclude that 
the de Vaucouleurs bulge with exponential disk profile is the most appropriate
way to investigate the structural properties. As our sample comprises only
early-type galaxies we use a similar approach, namely a
combination of an $r^{1/4}$ plus exponential disk profile by 
\citeauthor{Sag:97a} (\citeyear{Sag:97a}). These authors demonstrated that the
S\'{e}rsic $r^{1/n}$ profiles can be understood as a ``subset'' of
$r^{1/4}$+exponential models.
This algorithm also explores the effect of an efficient sky-subtraction.
For example, for $n>4$ large extrapolations are needed which could lead to
large errors in the sky-subtraction of up to $\pm3$\% and, therefore, to
additional uncertainties in the absolute magnitudes and half-light-radii.

\subsubsection{The surface brightness models}

For analysing structural parameters we use the fitting algorithm developed by
\citeauthor{Sag:97a} (\citeyear{Sag:97a},~\citeyear{Sag:97b}). The galaxies'
parameters are derived searching for the best combination of seeing-convolved,
sky-corrected $r^{1/4}$ and exponential laws. This approach
accounts for various types of observed luminosity profiles, i.e. it models the
extended luminosity profiles of cD galaxies and provides fits to the range of
profile shapes of early-type galaxies (ellipticals featuring a flat core to
S0 galaxies with the presence of a prominent disk).

Subframes of all 14 early-type galaxies both with spec\-tro\-scopic and \hst\
information (without the cD galaxy) were extracted and each galaxy was analysed 
individually. In a first step stars and artifacts around the galaxies were
masked in order not to cause any problems in the fitting process. In the next
step the circularly averaged surface brightness profile of the galaxy was
fitted with PSF-convolved $r^{1/4}$ and exponential components, both
simultaneously and separately (see also Z01
and \citeauthor{Fri:04} \citeyear{Fri:04}).
This method allowed to derive the effective (half-light)
radius $R_{e}$ (in arcsec), the total $F814W$-band magnitude $I_{814}$
and the mean surface brightness within $R_{e}$, $\langle\mu_e\rangle$,
for the entire galaxy as well as the luminosity and scale of the bulge
($m_{b}$ and $R_{e,b}$) and disk ($m_{d}$ and $h$) component separately,
within the limitations described by \cite{Sag:97a}.

In total, structural parameters could be determined for 14 galaxies out of 15
for which we also have obtained spectra (see Table~\ref{hstpar}). In
Fig.~\ref{SBPHSTgals} of the Appendix examples of the surface brightness
profile fits are shown. In most cases, the observed profile is best described
by a combination of a bulge (de Vaucouleurs law) and a disk component
(exponential law). Typical residuals $\Delta \mu_{I}$ between the observed
profile and the fit as a function of $R^{1/4}$ are in the
order of $0.01^{\rm m}\leq\Delta \mu_{I}\leq 0.10^{\rm m}$.
Comparing the differences between elliptical and lenticular
(S0) galaxies we detect a small disk component for the S0 exhibiting a
lens-like structure. For example, the elliptical galaxy
\#\,2438 is well represented by a pure de Vaucouleurs profile. In the case of
the S0 galaxy \#\,2946 a combined model of an
$r^{1/4}$-law and an additional disk component results in the best luminosity
profile. For the innermost regions ($R^{1/4}\la 0.5$), the model is
extrapolated to the calculated central surface brightness $\mu_{0}$
of the galaxy.

For some cases (\#\,2198, \#\,2438 and \#\,2763) we do not detect any additional
disk component, i.e. a pure classical de Vaucouleurs law profile ($D/B=0$)
results in the best profile for the galaxy's light distribtuion. Therefore,
our approach also allows for the possibility that dynamically hot galaxies may
all have $r^{1/4}$ profiles. But most of our galaxies are well
described by the superposition of $r^{1/4}$ bulge and exponential disk profiles.
Bulge+disk models provide better fits than a pure $r^{1/4}$ law, and
the addition of an exponential disk component improves the fit for many of the
galaxies. The surface brightness at the half-light radius is a strong function
of the chosen fitting profile and the half-light radius is dependent
on the assumed surface brightness model. However, the product
$R_{{\rm e}}\,\langle I_{{\rm e}}\rangle^{\beta}$ (with $\beta\simeq0.8$)
provides a Fundamental Plane parameter which is stable for
\textit{all} applied profile types, from $r^{1/4}$ to exponential laws
(\citeauthor{SBD93} \citeyear{SBD93}; \citeauthor{KIDF00a} \citeyear{KIDF00a}).
Under the constraint of this coupling, galaxies can only move parallel to the
edge-on view of the Fundamental Plane.

Some previous Fundamental Plane studies derived their structural parameters
with a pure $r^{1/4}$ profile. Therefore, we have also tested our results using
surface brightness models based on a pure de Vaucouleurs law only.
Only systems featuring a dominant disk (Sa bulges) show some deviations.
As expected, these galaxies move only along the edge-on projection of the
Fundamental Plane. Note, that the errors of the
FP parameters of effective radius ($R_{{\rm e}}$) and surface brightness
($\mu_{{\rm e}}$) are highly correlated. However, the rms uncertainty in the
product $R_{{\rm e}}\,I_{{\rm e}}^{0.8}$ which propagates into the FP is
only $\approx$4\,\%, corresponding to
$\Delta \mu_e=\beta\,\Delta\,{\rm log}\,R_e$ with $\beta=0.328$ and a scatter
of the latter relation of 0.05~mag arcsec$^{-2}$
(\citeauthor{SBD93} \citeyear{SBD93}). The results of the $r^{1/4}$ structural
measurements are presented in Table~\ref{dVhstpar}. Fig.~\ref{FPedge} shows the
Fundamental Plane for A\,2218 and A\,2390 constructed based on a pure
$r^{1/4}$-law profile. The $r^{1/4}$-law structural parameters for A\,2218 can
be requested from the first author.

{\scriptsize
\begin{table*}
\centering
\begin{minipage}{160mm}
\caption{Galaxy properties of the \hst\ sub-sample. (1) galaxy ID,
(2) velocity dispersion $\sigma_{c}$ (aperture corrected) with errors [in \kms],
(3) extinction-corrected, Vega-based total magnitude in WFPC2 F814W filter,
(4) mean surface brightness magnitude $\langle \mu_{e}\rangle$ in Gunn $r$
corrected for cosmic expansion, (5) rest-frame Gunn $r$ magnitude 
[in mag], (6) effective radius [in arcsec] and (7) effective radius [in kpc],
(8) effective radius of the bulge [in arcsec],
(9) FP parameter $R_{{\rm e}}\,I_{e}^{0.8}$, (10) disk scale-length [in arcsec],
(11) disk-to-bulge ratio ($D/B$), (12) morphology of the object.}
\label{hstpar}
\begin{tabular}{rcccccrccccc}
\hline\hline
\noalign{\smallskip}
ID & $\sigma_{c}$ & $I_{{\rm 814}}$ & $\langle \mu_{e}\rangle_{r}$ &
$M_{r}$ & $R_{{\rm e}}$ & log $R_{{\rm e}}$ &
$R_{{\rm e,b}}$ & $R_{{\rm e}}I_{e}^{0.8}$ & h & $D/B$ & morp \\
 & [\kms] & [mag] & [mag/arcsec$^{2}$] & [mag] & [arcsec] & [kpc] & [arcsec] & &
 [arcsec] & & \\
 (1) & (2) & (3) & (4) & (5) & (6) & (7) & (8) & (9) & (10) & (11) & (12) \\
\noalign{\smallskip}
\hline
\noalign{\smallskip}
2106 & 144.5$\pm 16.8$ & 19.18 & 18.28 & 19.93 & 0.282 & 0.013 & 0.250 & 0.033 & 0.20 & 0.36 & E    \\
2120 & 140.3$\pm 15.3$ & 18.88 & 18.22 & 19.63 & 0.314 & 0.060 & 0.202 & 0.154 & 0.34 & 0.46 & S0   \\
2138 & 165.4$\pm 08.7$ & 16.97 & 20.60 & 17.72 & 2.264 & 0.918 & 1.324 & 1.800 & 4.48 & 0.36 & E    \\
2180 & 161.2$\pm 10.6$ & 17.80 & 19.49 & 18.55 & 0.928 & 0.530 & 0.580 & 1.195 & 0.88 & 0.60 & S0   \\
2198 & 148.9$\pm 15.6$ & 18.59 & 17.92 & 19.34 & 0.313 & 0.058 & 0.313 & 0.155 & 0.00 & 0.00 & E    \\
2237 & 162.2$\pm 12.3$ & 17.66 & 19.94 & 18.41 & 1.217 & 0.648 & 1.195 & 1.385 & 0.73 & 0.81 & Sa   \\
2438 & 119.8$\pm 15.9$ & 18.74 & 19.59 & 19.49 & 0.631 & 0.363 & 0.631 & 0.808 & 0.00 & 0.00 & E    \\
2460 & 258.6$\pm 10.2$ & 18.00 & 18.90 & 18.75 & 0.642 & 0.370 & 0.509 & 0.892 & 1.09 & 0.17 & E    \\
2592 & 192.0$\pm 11.4$ & 17.20 & 20.11 & 17.95 & 1.625 & 0.774 & 1.104 & 1.618 & 2.55 & 0.29 & E    \\
2619 & 229.3$\pm 14.3$ & 18.53 & 18.20 & 19.28 & 0.366 & 0.126 & 0.320 & 0.326 & 0.27 & 0.37 & E    \\
2626 & 195.6$\pm 12.7$ & 18.31 & 18.65 & 19.06 & 0.497 & 0.259 & 0.325 & 0.640 & 0.38 & 0.92 & S0   \\
2763 & 300.5$\pm 12.2$ & 17.55 & 18.89 & 18.30 & 0.789 & 0.460 & 0.789 & 1.108 & 0.00 & 0.00 & E    \\
2946 & 131.9$\pm 15.2$ & 18.69 & 19.26 & 19.44 & 0.554 & 0.306 & 0.405 & 0.708 & 0.60 & 0.34 & S0   \\
6666 & 211.0$\pm 13.5$ & 17.65 & 18.79 & 18.40 & 0.721 & 0.421 & 0.488 & 1.024 & 1.00 & 0.32 & E/S0 \\
\noalign{\smallskip}
\noalign{\hrule}
\end{tabular}
\end{minipage}
\end{table*}}

{\scriptsize
\begin{table*}
\centering
\begin{minipage}{160mm}
\caption{Galaxy parameters of the \hst\ sub-sample based on pure
de Vaucouleurs surface brightness fits. Columns are analogous tabulated as in
Table~\ref{hstpar}. The velocity dispersions $\sigma$ with corresponding
errors [in \kms] are not aperture corrected.}
\label{dVhstpar}
\begin{tabular}{rcccccrcc}
\hline\hline
\noalign{\smallskip}
ID & $\sigma$ & $I_{{\rm 814}}$ & $\langle \mu_{e}\rangle_{r}$ &
$M_{r}$ & $R_{{\rm e}}$ & log $R_{{\rm e}}$ & $R_{{\rm e,b}}$
& $R_{{\rm e}}I_{e}^{0.8}$ \\
 & [\kms] & [mag] & [mag/arcsec$^{2}$] & [mag] & [arcsec] & [kpc] & [arcsec] & \\
 (1) & (2) & (3) & (4) & (5) & (6) & (7) & (8) & (9) \\
\noalign{\smallskip}
\hline
\noalign{\smallskip}
2106 & 131.0$\pm 16.8$ & 19.07 & 18.55 & 19.82 & 0.335 & 0.088 & 0.335 & 0.219 \\
2120 & 127.2$\pm 15.3$ & 18.75 & 18.51 & 19.50 & 0.382 & 0.145 & 0.382 & 0.363 \\
2138 & 150.0$\pm 08.7$ & 17.14 & 20.04 & 17.89 & 1.621 & 0.772 & 1.621 & 1.630 \\
2180 & 146.2$\pm 10.6$ & 17.67 & 19.79 & 18.42 & 1.133 & 0.617 & 1.133 & 1.342 \\
2198 & 135.0$\pm 15.6$ & 18.59 & 17.92 & 19.34 & 0.313 & 0.058 & 0.313 & 0.155 \\
2237 & 147.1$\pm 12.3$ & 17.12 & 21.10 & 17.87 & 2.667 & 0.989 & 2.667 & 1.802 \\
2438 & 108.6$\pm 15.9$ & 18.74 & 19.59 & 19.49 & 0.631 & 0.363 & 0.631 & 0.808 \\
2460 & 234.5$\pm 10.2$ & 18.01 & 18.86 & 18.76 & 0.628 & 0.361 & 0.628 & 0.872 \\
2592 & 174.1$\pm 11.4$ & 17.27 & 19.87 & 18.02 & 1.408 & 0.711 & 1.408 & 1.533 \\
2619 & 207.9$\pm 14.3$ & 18.40 & 18.55 & 19.15 & 0.456 & 0.222 & 0.456 & 0.553 \\
2626 & 177.4$\pm 12.7$ & 18.01 & 19.29 & 18.76 & 0.768 & 0.448 & 0.768 & 1.033 \\
2763 & 272.5$\pm 12.2$ & 17.55 & 18.89 & 18.30 & 0.789 & 0.460 & 0.789 & 1.108 \\
2946 & 119.6$\pm 15.2$ & 18.62 & 19.36 & 19.37 & 0.599 & 0.340 & 0.599 & 0.778 \\
6666 & 191.3$\pm 13.5$ & 17.65 & 18.74 & 18.40 & 0.701 & 0.408 & 0.701 & 1.000 \\
\noalign{\smallskip}
\noalign{\hrule}
\end{tabular}
\end{minipage}
\end{table*}}

Thumbnail images for the 14 galaxies are provided in the Appendix in
Fig.~\ref{hstthumbs}. A detailed description of the morphological classification
of galaxies residing in the \emph{HST} field is outlined in the Appendix
\S\ref{hstprop}. The final classification is listed in Table~\ref{hstpar} and
is also given for each galaxy in Fig.~\ref{hstthumbs}.

As an additional comparison, we also performed an isophote analysis based on the
procedure introduced by \cite{BM87}. Deviations from the elliptical isophotes
were recorded as a function of radius by a Fourier decomposition
algorithm. The presence and strength of the $a_{4}$ coefficient, which
represents the signature of diskyness, is in good agreement with the visually
classified morphologies of S0 and spiral bulges.

\subsubsection{Error evaluation}\label{erreva}

The errors of the photometric parameters of effective radius and surface
brightness are correlated and enter the FP in combination. 
The errors in $R_{e}$ (in arcseconds) and $\mu_{e}$
(in magnitudes) propagate into the FP relationship in the following form
(\citeauthor{Treu01a} \citeyear{Treu01a}):
\begin{equation}
\delta FP_{\rm phot} = {\rm log}\ \delta R_{e}-\beta\ \delta \mu_{e} \label{fperr}
\end{equation}
An error calculation with this formula is not an approximation but a
particularly robust computation, as in three-dimensional space defined by the
FP it is the only natural way to provide an accurate error estimation (see also
\citeauthor{Sag:97a} \citeyear{Sag:97a}; \citeauthor{KIDF00a} \citeyear{KIDF00a};
\citeauthor{Treu01a} \citeyear{Treu01a} for further details).
For this reason, we show the errors in the FP only in the edge-on projection
along the short axis that separates kinematic measurements and photometric
parameters (see upper right panel in Fig.~\ref{FPr}). The errors are calculated
for the F814W filter with a parameter value of $\beta=0.328$. Since the slope
$\beta$ is very well defined with only a small variation with wavelength, the
error estimations change negligibly within the observed range of $\beta$ values
(\citeauthor{PDC98a} \citeyear{PDC98a}).

The errors on the derived structural parameters are assigned through a
combination of individual quality parameters, including the effects of sky
subtraction errors, seeing and pixel sampling, average galaxy surface
brightness relative to the sky, $S/N$ variations, extrapolation involved to
derive $M_{{\rm tot}}$, radial extent of the profile and reduced $\chi^{2}$ of
the fit resulting in a final global quality parameter.
The total quality parameter $Q$ is defined by the maximum of the quality
parameters of individual errors $Q$ = Max(Q$_{{\rm max}}$, Q$_{\Gamma}$,
Q$_{S/N}$, Q$_{{\rm Sky}}$, Q$_{\delta {\rm Sky}}$, Q$_{{\rm E}}$,
Q$_{\chi^{2}}$) (\citeauthor{Sag:97a} \citeyear{Sag:97a}).
For $Q=1$ the typical uncertainty in total magnitude is
$\Delta M_{{\rm tot}}=0.05$ and in effective radius
$\Delta$~log~$R_{e}=0.04$ ($<10$\%). A value of $Q=2$ leads to errors of
$\Delta M_{{\rm tot}}=0.15$ and $\Delta$~log~$R_{e}=0.1$ ($<25$\%),
respectively. A quality parameter of $Q=3$ ensues errors of
$\Delta M_{{\rm tot}}=0.4$ and $\Delta$~log~$R_{e}=0.3$.
For A\,2390 we find that 43\% (6) of our galaxies have $Q=1$,
57\% (8) galaxies have $Q=2$ and no galaxy yield a quality parameter
of $Q=3$. We estimate the average error in $I_{814}$ to be $\approx$10\,\% 
and the error in $R_{e}$ of $<25$\,\%. 

Furthermore, we include variations in the zero-point calibration
$\delta {\rm ZP}=0.03$ (\citeauthor{Sag:97b} \citeyear{Sag:97b}), uncertainty
in galactic extinction $\delta {\rm E(B-V)}=0.010$, as well as errors
in the k-correction of $\delta m_{r}=0.02^{\rm m}$.
The combination of the uncertainties on the measured parameters
$\delta \mu_{e}$ [in mag] and ${\rm log}\ \delta R_{e}$ [in arcsec] according
to equation~\ref{fperr} are listed as $\delta {\rm FP_{\rm phot}}$. Adding the
errors of galactic extinction correction, the k-correction and the errors in
zero-point in quadrature ensues the final total error of the FP in the
rest-frame Gunn $r$-band, $\delta {\rm FP_{r}}$. 
A summary of our error analysis is given in Table~\ref{errana}.

{\scriptsize
\begin{table}
\centering
\begin{minipage}{0.45\textwidth}
\caption{Error analysis for the FP. The total error on the combination that
enters the FP,
$\delta FP_{\rm phot} = {\rm log}\ \delta R_{e}-\beta\ \delta \mu_{e}$
(see equation~\ref{fperr}), is listed as $\delta {\rm FP_{\rm phot}}$, adopting
$\beta=0.328$.  The final FP error in the rest-frame properties is given in the
column $\delta {\rm FP}_{r}$.}
\label{errana}
\begin{tabular}{ccrcc}
\hline\hline
\noalign{\smallskip}
ID & $\delta\mu_{e}$ & ${\rm log}\ \delta R_{e}$ &
$\delta {\rm FP_{\rm phot}}$ & $\delta {\rm FP}_{r}$\\
\noalign{\smallskip}
\hline
\noalign{\smallskip}
 2106 & 0.078 & -0.055 & 0.081 & 0.082 \\
 2120 & 0.071 & -0.020 & 0.044 & 0.045 \\
 2138 & 0.136 &  0.124 & 0.080 & 0.081 \\
 2180 & 0.049 & -0.001 & 0.017 & 0.019 \\
 2198 & 0.073 & -0.020 & 0.044 & 0.046 \\
 2237 & 0.076 &  0.050 & 0.025 & 0.027 \\
 2438 & 0.027 & -0.040 & 0.049 & 0.051 \\
 2460 & 0.030 & -0.019 & 0.029 & 0.031 \\
 2592 & 0.105 &  0.094 & 0.059 & 0.061 \\
 2619 & 0.054 & -0.018 & 0.035 & 0.037 \\
 2626 & 0.029 & -0.030 & 0.040 & 0.041 \\
 2763 & 0.032 & -0.004 & 0.015 & 0.016 \\
 2946 & 0.021 & -0.026 & 0.033 & 0.034 \\
 6666 & 0.028 & -0.014 & 0.023 & 0.025 \\
\noalign{\smallskip}
\noalign{\hrule}
\end{tabular}
\end{minipage}
\end{table}
}

\subsubsection{Determination of rest-frame surface brightnesses}
\label{rfsb}

At the cluster redshift of A\,2390, the observed F814W passband is
close to rest-frame Gunn $r_{{\rm rest}}$. Therefore, it is more promising to
use a single k-correction term of the observed $I_{814}$ to Gunn $r$ in
rest-frame rather than to take a more complicated way in converting to $R_{c}$
first and then using a colour transformation. The latter procedure would
increase the overall uncertainties.

Analysis of the rest-frame luminosities of the galaxies, was performed
in the following way. According to \cite{Hol:95}, the instrumental magnitude
in the F814W filter is given by:
\begin{equation}
I_{814} = -2.5 \log({\rm DN})/t_{{\rm exp}}+ZP+2.5\log(GR)
\end{equation}
with $GR$ being the respective gain ratios for the WF chips and $ZP$ the
zero-point for an exposure time $t_{{\rm exp}}=1$\,s. $ZP=20.986$ with
consideration of the difference between ``short'' and ``long'' exposures of 
0.05\,mag (\citeauthor{Hil:98} \citeyear{Hil:98}) and an aperture correction
of 0.0983 (\citeauthor{Hol:95} \citeyear{Hol:95}).

Since we do not have any ground-based $V-I$ colours a similar transformation
from Cousins $I$ to Gunn $r$ following the procedure as in Z01
can not be conducted. Instead, we used the spectral template for a
typical E/S0 galaxy by \cite{KCBMSS96}, redshifting it to the
cluster redshift of $z=0.23$, to determine the flux through the
$I_{814}$ filter and at $z=0$ through the \cite{TG76} $r$ filter. This lead to
the translation $I_{814\ {\rm obs}}-{\rm Gunn}\ r_{{\rm rest}}=-0.75$. We
checked this result by transforming the synthetic SEDs by \cite{Moe01},
yielding only a slightly higher value of $0.02$~mag.

The mean surface brightness within $R_e$ is defined as:
\begin{equation}
\langle\mu_{r}\rangle_{e} = r_{{\rm rest}}+2.5 \log (2\pi)+5 \log(R_{e})-10 \log(1+z),
\end{equation}
where the parameter $r_{{\rm rest}}$ denotes rest frame Gunn $r$ magnitude.
The dimming due to the expansion of the Universe is corrected by the last term
of this equation. Applying the formula in Z01, the mean
surface brightness $\langle I\rangle_{e}$ in units of L$_{\sun}\,$/pc$^{2}$
was calculated. Furthermore, the effective radius in kpc was computed for our
cosmology with an angular distance of A\,2390 of $d=753.41$\,Mpc. For Coma
($z=0.024$) the angular distance is $d=99.90$\,Mpc.


\section{Spectroscopy of Abell~2390}\label{spec}

\subsection{Sample selection}

\begin{figure*}
\vspace*{0.3cm}
\centerline{%
\includegraphics[width=1.0\textwidth]{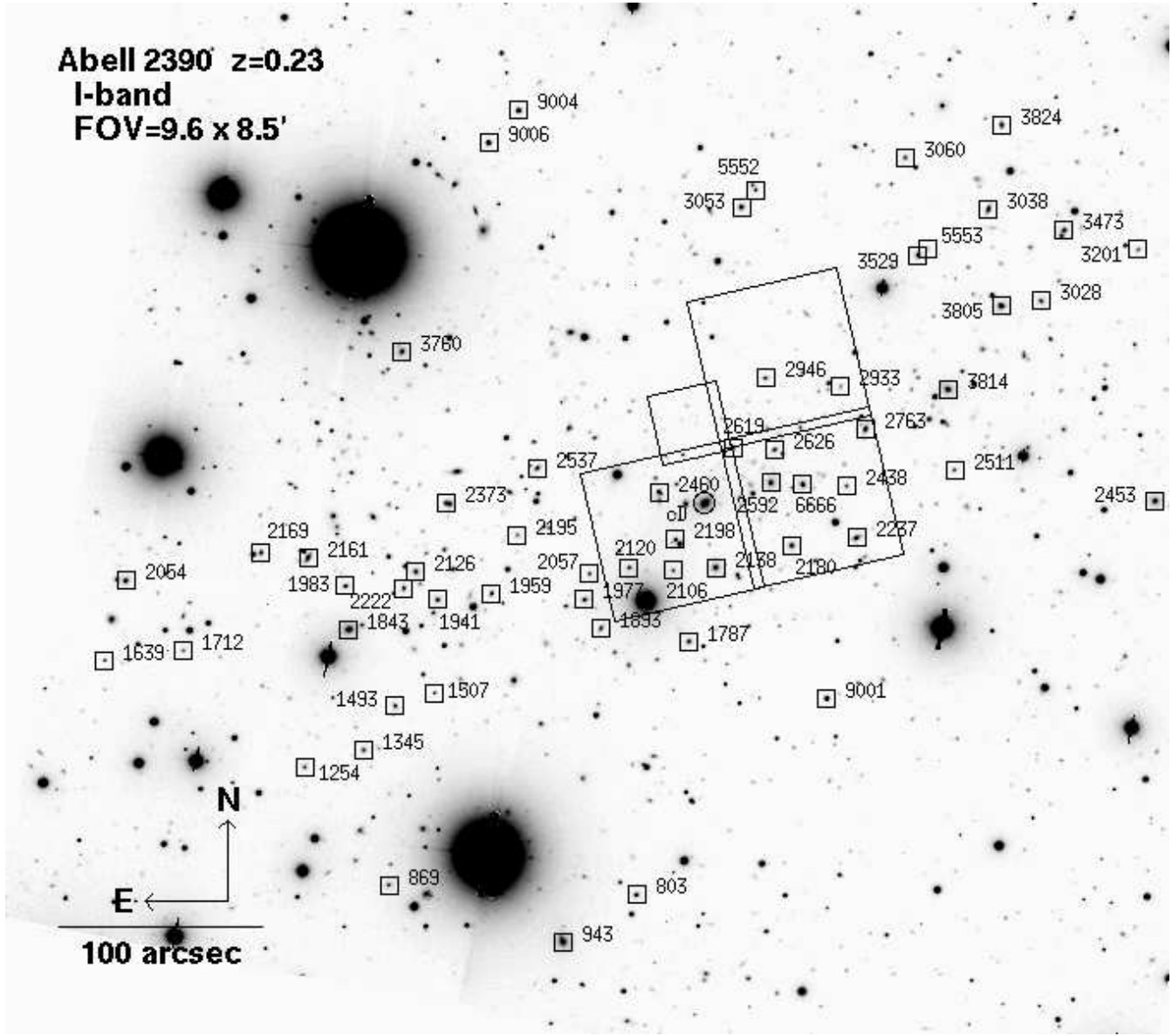}}
\caption{$I$-band image of the cluster Abell~2390 at $z=0.23$, taken with
the 5.1-m Hale telescope on Mt. Palomar. The total FOV is
$9\farcm6\times8\farcm5$. North is up, east to the left. The length of the
scale bar in the left corner corresponds to $100''$, or 365\,kpc at the
distance of A\,2390 ($\Omega_{m}=0.3$, $\Omega_{\Lambda}=0.7$, $h^{-1}_{70}$).
Galaxies with spectra are indicated by squares.
The cD galaxy is marked with a circle. Stars used for the mask alignment are
labelled as ${\rm ID}=900{\rm x}$. An overlay of the HST WFPC2 FOV is also
indicated.}
\label{HALE}
\end{figure*}

In order to enable a good sky subtraction which requires long slit lengths
(with a minimum length of 15$''$), each mask was constrained to only
about 20 galaxies in total. For this reason, we were very careful to select
only galaxies which were likely to be cluster members based upon their
$UBI$ broad-band colours. The target objects were selected on the basis of
the ground-based $I$-band images and a combination of defined colour
regions. Using a similar selection procedure as described in Z01,
all galaxies were rejected which were falling outside a 
defined colour range, restricted to $2.60<(B-I)<3.50$ and $-0.2<(U-B)<0.40$.
This method puts negligible restrictions on the stellar populations of the
selected objects, but eliminates the majority of background galaxies.
Since our galaxies were distributed over the whole field-of-view (FOV) of
$\sim$10$'\times 10'$ (see Fig.~\ref{HALE}), corresponding to
$1.53\times 1.53\ h^{-2}_{70}$~Mpc$^{2}$, we can study the evolution of
early-type galaxies out to large clustercentric distances.

\subsection{Observations}
\label{obs}

The observations were carried out during two observing runs in four nights of
September 7.--11., 1999 and five nights of July 26.--31., 2000. To achieve
intermediate resolution spectra for the galaxies in A\,2390 at $z=0.23$, we
chose the grism {\sc green 1000}
with a typical wavelength range of $\lambda\lambda=4500$--7500\AA, encompassing
important absorption lines such as \hg, \hb, \mgb, Fe5270 and Fe5335 at
the cluster redshift of $z=0.228$. The slit widths were set to 1.5\arcsec
and the instrumental resolution
around \hb\ and \mgb\ ($5900\la\lambda_{{\rm obs}}\la 6400$~\AA) was
5.5~\AA\ FWHM, corresponding to $\sigma_{\rm inst}\sim 100$~km~s$^{-1}$. 
The spatial scale was 0.33\arcsec\ pixel$^{-1}$ and seeing conditions
varied between $1.2\arcsec\le{\rm FWHM}\le 1.7\arcsec$.

Three masks were observed with total exposure times between 29880~sec (8.3 hrs)
and 42480~sec (11.8 hrs) each (see Table~\ref{tab-obs}).
In total, we obtained 63 high-signal-to noise spectra of 52 different galaxies,
of which 15 are situated within the {\it HST} field. 
Three objects, \#\,1507, \#\,1639 and \#\,2933, turned out to be background
galaxies at $z=0.3275$, $z=0.3249$ and $z=0.3981$, respectively; one galaxy
(\#\,3038) is located in the foreground at $z=0.1798$. We discard these
galaxies from our sample yielding a total of 48 different early-type cluster
members, proving that our sample selection was highly efficient.

\begin{figure}
\centerline{\psfig{figure=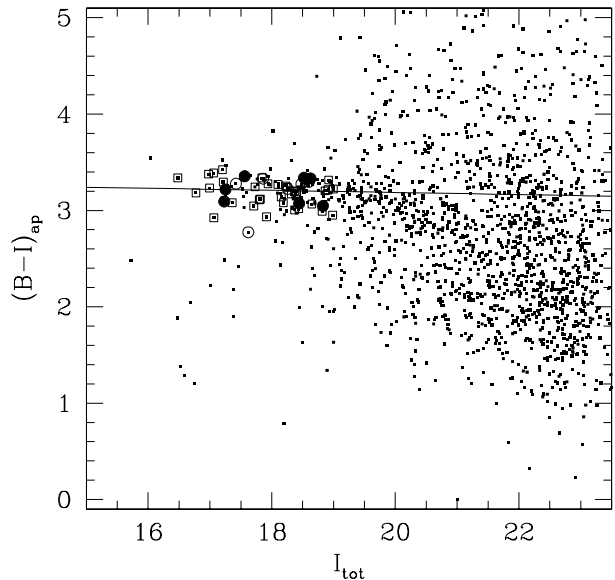,width=80mm}}
\caption{$(B-I)-I$ colour-magnitude diagram for galaxies brighter than
$I=23.5$~mag, obtained from our HALE imaging and the 48 observed early-type
cluster members of A\,2390 with available spectra (squares).
Circles represent those galaxies
in A\,2390 which enter the Fundamental Plane, ellipticals (filled circles),
S0 galaxies (open circles). The solid line is a least-square fit to seven
ellipticals.}
\label{cmdbi}
\end{figure}

In Fig.~\ref{cmdbi} the colour-magnitude diagram $(B-I)$ versus $I$ for
our 48 early-type cluster members of A\,2390 with available spectroscopic
information is shown. A least-squares fit to the seven ellipticals in
A\,2390 gives
$(B-I)_{{\rm E}}=-0.011\ (I_{{\rm tot}})+3.405$ which is indicated by the
solid line in Fig.~\ref{cmdbi}. The outlier object \#\,2237
with the bluest colour of $(B-I)=2.77$ was classified a spiral Sa galaxy.


\subsection{Data reduction}
\label{specredu}

The reduction of the spectra was undertaken using
{\sc midas}\footnote{ESO--MIDAS, the European Southern 
Observatory Munich Image Data Analysis System is developed and maintained by 
the European Southern Observatory.} with own {\sc
Fortran} routines and followed the standard procedure.
For each observing run, bias frames were taken at the beginning and end
of the night. All frames showed a very stable two--dimensional structure, thus
all bias frames from the individual nights were used to generate master biases,
one for each observing run.
An averaged super bias image was constructed of median--scaled individual bias
frames. After calculating the median in the overscan region of the super bias,
the super bias frame was scaled to median of each individual bias image. Finally
this scaled master bias frame was subtracted from each frame. Variations in the
bias level were $\leq$5\%.

The 2-d images of each slit spectrum were extracted from the MOS frame
(after bias subtraction) and reduced individually. Special care was taken in
correcting for the S--distortions (curvature) of the spectra, which were
strongest when lying closest to the edges of the field. For this purpose a
special IRAF program was used which fits a user-defined
Legendre polynomial (in most cases of third order) to compute the parameters
for the curvature correction. The same set of coefficients were applied to the
science, flatfield and corresponding wavelength calibration frames,
respectively.

A minimum number of five dome flat--fields were used to construct a master
flatfield through a combination of median--averaged single flatfield exposures.
After the rectification, the flatfield slit exposures were approximated by a
spline function of third degree to the continuum of each spectrum
(in dispersion direction) to account for the CCD response curve.
The smoothed spline fit was afterwards applied for normalisation of the
original flatfield frame in order to correct for pixel-to-pixel variations.

Cosmic ray events were removed very careful by a $\kappa$--$\sigma$ clipping
algorithm with a $5\times 5$ pixel filter, taking into account only to
eliminate possible artefacts in the spectra. Furthermore, up to five bad
columns per slitlet were cleaned by interpolating from unaffected adjacent
columns.

The sky background was subtracted by iteratively fitting each CCD column
separately through a $\kappa$--$\sigma$ clipping algorithm.
To ensure consistent profile centers to within at least half a pixel,
corresponding to 0.17\arcsec, the one-dimensional spectra were extracted
for each exposure separately using the Horne--algorithm \citep{Horne86}, which
optimally weights the extracted profile to maximise the signal-to-noise.
Thus, possible shifts of the spectra which may
have occurred during different nights are accounted for in the final
one-dimensional summed-up spectra.

\begin{figure*}
\vspace*{0.3cm}
\centerline{%
\includegraphics[width=0.80\textwidth]{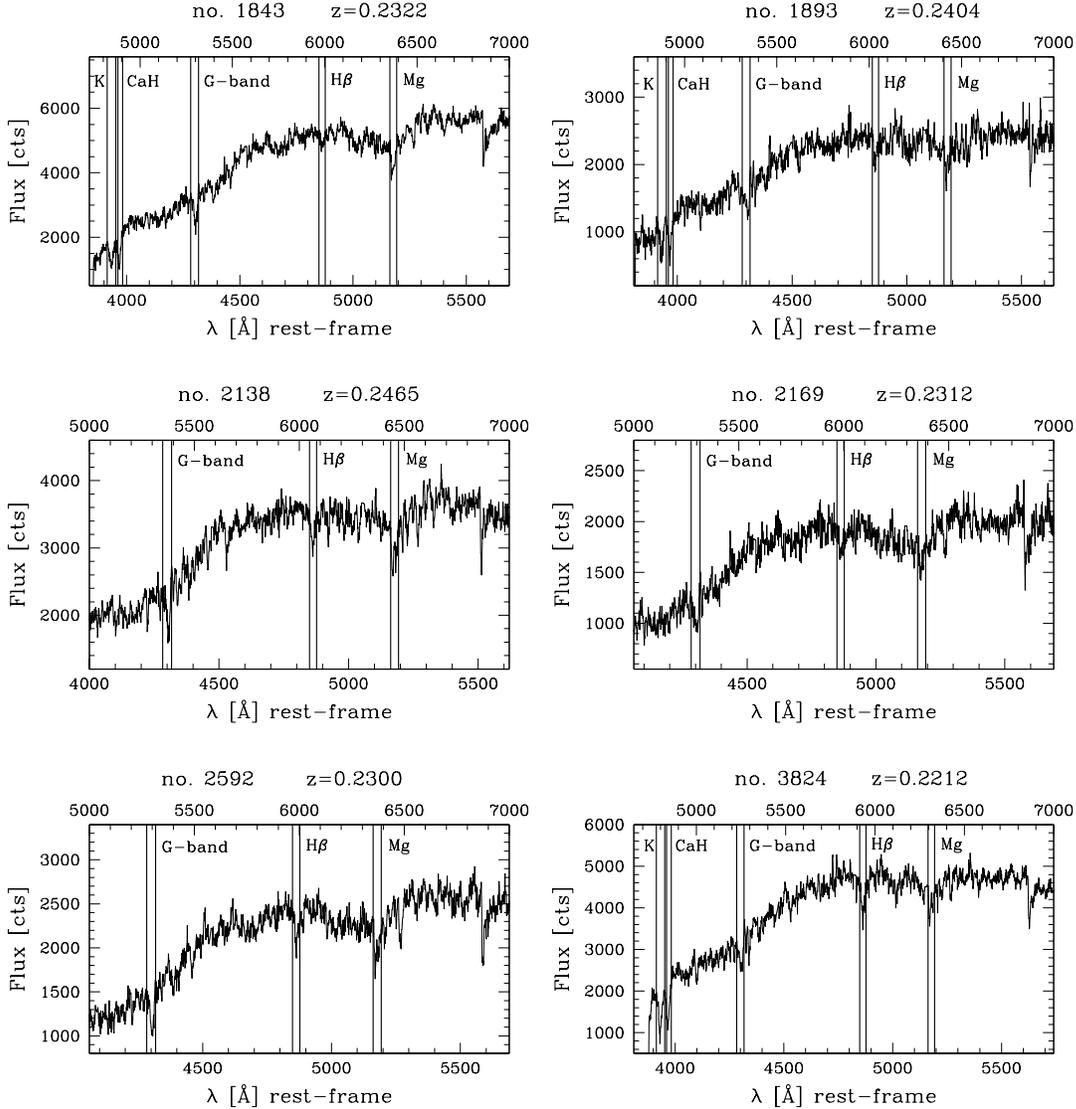}}
\caption{Example spectra (not flux-calibrated) of early--type galaxies
in our sample. The lower x-axis represents the rest frame wavelengths, the
upper one the observed wavelengths (both in \AA). The ordinate gives the flux
in counts (1~ADU=1.1~$e^{-1}$). Prominent absorption features are marked and
the ID and the determined redshift are given above each panel.}
\label{sperestfr}
\end{figure*}

For the wavelength calibration exposures, the HgAr and Ne lamps were 
switched on together in order to gain a sufficient number of emission lines
over a large range, $\lambda\lambda\approx 5450$--7450\,\AA. A two--dimensional
dispersion relation was computed using polynomial fit functions of third and
second degree in dispersion and in spatial direction, respectively. Typical
r.m.s. errors of the dispersion fit were 0.05-0.06\,\AA\ at a stepsize of
1.3\,\AA\ per pixel. After the wavelength calibration, individual exposures
were summed up and the final one-dimensional spectrum (see Fig.~\ref{sperestfr})
was rebinned to logarithmic wavelength steps in preparation for the
determination of the velocity dispersion using the
{\it Fourier Correlation Quotient} (FCQ) method
(\citeauthor{Ben:90} \citeyear{Ben:90}) and measurement of absorption
line strengths.

In a similar manner, the spectra of standard stars were reduced.
Two spectrophotometric flux standards (HZ2 and HZ4) were observed through
an acquisition star hole in one mask. For the kinematic templates, three
K giant stars (SAO\,32042 (K3III), SAO\,80333 (K0III), SAO\,98087 (K0III))
were observed through a 0.5\arcsec longslit using the same grism as for the
galaxies. These stars had a spectral resolution
at \hb\ and \mgb\ of $\sim$2.2\,\AA\ FWHM, corresponding to
$\sigma_{*}=55$~\kms). To minimise the effect of any possible variation in slit
width, the star spectra were averaged over a small number of rows only.

\subsubsection{Velocity dispersions}

Galaxy velocity dispersions and radial velocities were calculated
using an updated version of the FCQ program kindly provided
by Prof.\ R.\ Bender (\citeauthor{Ben:90} \citeyear{Ben:90}).
At a redshift of $z=0.231$ the \mgb\ line is strongly
affected by the telluric emission line at $\lambda=6367$\,\AA.
For this reason, two separate wavelength regimes were
analysed. The first wavelength range between $\lambda\lambda=6204$--6748\,\AA\
was centered on the \mgb\ feature. In those cases where the \mgb\ line was
affected, either in the central or in the adjacent continuum windows, a reliable
measurement of the velocity dispersion could not be ensured so that a second
calculation in the range around the G4300 band between
$\lambda\lambda=4964$--5541\,\AA\ was performed.

Since the slitlets of the galaxies had small variations in width size and were
not identical in width to the longslit of the stars, a simple
comparison and interpretation of the calculated velocity dispersions would be
wrong. Thus, we applied a correction procedure to the resulting dispersions to
correct for this effect. In a first step, the instrumental dispersion was
determined by measuring the width of four unblended emission lines around the
position of the redshifted \mgb\ in the respective arc spectrum of a galaxy.
Typical instrumental resolutions are in the range of
$\sigma_{\rm inst}\sim 90-100$~km~s$^{-1}$. The spectra of the stars
were broadened to the same resolution (in \kms) as the galaxies, before using
them as templates in the FCQ algorithm to compute the velocity dispersion for
each galaxy. This procedure has the advantage that the galaxy velocity
dispersion results directly and no additional correction for the velocity
dispersion of the template star $\sigma_{\rm star}$ is needed (see Z01
for an alternative approach). All values
of $\sigma_{{\rm gal}}$ are given in Table~\ref{photprop} together with their
corresponding heliocentric radial velocities $v$.

For a total of eight galaxies which are included in two different MOS masks,
we are able to confirm the internal reliability of our
spectral analysis. The velocity dispersions have been measured for both
individual setups and the agreement between these two determinations is
good (with a median offset for $\sigma$ at the 10\,\% level). In order
to increase the signal-to-noise ($S/N$) in our final spectra, these
repeat observations have been co-added. The $S/N$ in the final spectra varies
between 21 and 65 with a median value of $S/N\sim 34$ per \AA\ and a mean value
of $S/N\sim 37$ per \AA.

Velocity dispersions were aperture corrected according to the method
outlined by \cite{MTS03}. This aperture correction is slightly steeper than
the one proposed by \cite{Joerg99} (but consistent within the errors). For our
$\sigma$ measurements this results in corrections for the cluster A\,2390 by
$\Delta ({\rm log}\,\sigma_{{\rm A\,2390}})=+0.042$ and the cluster A\,2218 by
$\Delta ({\rm log}\,\sigma_{{\rm A\,2218}})=+0.039$.
For A\,2390, the aperture radius $a$ was taken as the square root of the slit
width of $1.5''$ times the mean of number of rows over which the spectra
were summed up as $8.4$ pixels (corresponding to $2.8''$). More details on the
correction method are described in Z01.


\section{Comparison A\,2390 versus A\,2218}

In a recent study, Z01 performed a detailed analysis of a sample of 48
early-type galaxies in the rich cluster A\,2218 at a redshift of $z=0.18$.
Combining this work with our sample of $N=48$ in A\,2390 offers the possibility
to explore the evolution of $\sim$100 early-type galaxies over a large
luminosity range and wide field-of-view at similar cosmic epochs.
The combined sample allows an extensive investigation of the evolutionary
status of galaxies in rich clusters at $z\sim 0.2$. A sub-sample with
accurate structural parameters provided by \hst\ comprises 34 E+S0 galaxies,
splitted into 17 ellipticals (E), 2 E/S0, 9 S0, 3 SB0/a, 2 Sa bulges and
1 Sab spiral bulge that can be investigated in the FP.
With this large sample, possible radial and environmental dependences can be
explored in detail for the galaxy properties from the cluster centre to the
outskirts using the Faber-Jackson relation and for
different sub-populations (E/S0 and S0/Sa bulges).

Before the individual galaxies of the two clusters can be combined into one
large sample, it has to be proven that both data sets are characterised by 
very similar properties and that we accounted for possible differences within
the sub-samples. For this reason, we compare the cluster galaxies with respect
to their luminosity, colour, size and velocity dispersion distributions. Both
clusters feature similar global cluster properties (e.g., richness class,
$X$-ray luminosity). Furthermore, the sample selection and all observations,
especially the spectroscopy, have been carried out in very similar manner.
Therefore, we do not expect significant differences between our galaxy samples.

\begin{figure}
\centerline{\psfig{figure=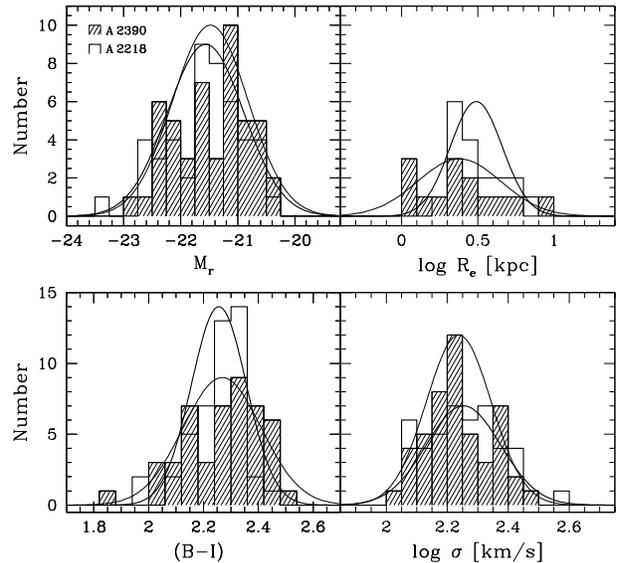,width=80mm}} 
\caption{Comparison of galaxy properties. Thin lines and hashed areas represent
the distribution for the members in A\,2390, solid thick lines the
characteristics of the galaxies in A\,2218. Gaussian fits showing the mean
values with $\pm1\sigma$ scatter are overlayed. {\it Top left:}\/ Absolute
Gunn $r$ magnitudes for the whole sample. {\it Top right:}\/ Size distribution
of the \hst\ sub-sample. {\it Bottom left:}\/ ($B-I$) rest-frame
colours of member galaxies. {\it Bottom right:}\/ Distribution of
velocity dispersions.}
\label{cluscomp}
\end{figure}

In the subsequent analysis the cluster galaxies are investigated over the same
dynamical range in their properties. Therefore, we excluded the two cD galaxies
of the cluster A\,2218 as no cD galaxy is included in the A\,2390 sub-sample.
A comparison of the cluster properties in absolute rest-frame Gunn $r$
magnitude, effective radius, ($B-I$) rest-frame colour and velocity dispersion
(aperture corrected) is presented in Fig.~\ref{cluscomp}. For each set of
parameters, the mean values with the resp. $\pm1\sigma$ scatter are indicated
as overlayed Gaussian curves. Overall, the galaxies are similarly distributed
in all plots.
For the galaxies in A\,2218 the range of absolute Gunn $r$ magnitudes
is $-20.50\ge M_{r}\ge-23.42$, for the objects in A\,2390 
$-20.47\ge M_{r}\ge-22.99$ (upper left panel in Fig.~\ref{cluscomp})
The median value for A\,2390 is $\langle {M}_{r}\rangle=-21.31^{m}$,
$0.17$~mag fainter than the median luminosity for the galaxies in A\,2218.
For the size distribution we consider only the 32 objects within
the \hst\ fields. In the distributions of galaxy size (upper right panel) small
(but not significant) differences are visible. The sizes of the A\,2218 galaxies
cover a range between 0.22 and 0.82 in ${\rm log}\,R_{e}$ (kpc), with a median
of ${\rm log}\,\langle{R_{e}}\rangle=0.46$. The A\,2390 sample
contains more galaxies with smaller effective radii
$0.01\ge {\rm log}\,R_{e}\ge0.92$, with a mean of
$\overline{{\rm log}\,R_{e}}=0.38\pm0.27$ (and a median of
${\rm log}\,\langle{R_{e}}\rangle=0.37$), resulting in a
broader distribution than for A\,2218. These galaxies are
low-luminosity galaxies. We will address this issue in more detail in the
forthcoming section~\S\ref{FP}. The rest-frame colour distributions show no
significant differences between the clusters. Using the predictions of passive
evolution models by \citeauthor{BC93} (\citeyear{BC93}; GISSEL96 version,
hereafter BC96), we corrected for the offset in $\Delta\,(B-I)$
between the two clusters, which is due to the difference in redshift.
The galaxies' colours in A\,2390 cover a range of $1.86<(B-I)<2.51$,
in A\,2218 $1.95<(B-I)<2.44$. For the A\,2390 galaxies we derive a median value
of $\langle(B-I)\rangle=2.29$, for the A\,2218 objects
$\langle(B-I)\rangle=2.28$, respectively. Both are in very good agreement with
the typical colour of $(B-I)=2.27$ for ellipticals at $z=0$, given
by \cite{FSI95}. The velocity dispersions for the galaxies are equally
distributed (lower right panel), with a median value of 165~\kms\ 
 ($\overline{{\rm log}\,(\sigma)}=2.238\pm0.11$) for A\,2390 and 178~\kms\
($\overline{{\rm log}\,(\sigma)}=2.253\pm0.12$) for A\,2218. As the velocity
dispersion is an indicator for the mass of an object and the measured $\sigma$
values exhibit similar ranges, we also conclude that there are no significant
differences in mean galaxy masses between the two samples
(see~\S\ref{mlr} for a further discussion). In addition,
we have also compared the scale lengths ($h$), disk-to-bulge ratios
($D/B$) and the surface brightnesses of our member galaxies and
again found negligible differences between the distributions.

As a conclusion we detect no significant offset in the distribution of any
galaxy parameters between the two samples. Therefore, we conclude that the
properties of the galaxies within the two clusters are very homogeneous and
thus a combination of the two data sets is adequate, resulting in a final
sample of 96 early-type galaxies.

%


\section{Results in A\,2390 and A\,2218}\label{res}

In all subsequent figures, large symbols denote the distant galaxies of
Abell~2218 and Abell~2390, small boxes represent the local reference sample.
Morphologically classified lenticular galaxies (S0) or bulges of early-type
spiral galaxies (from the A\,2218 and A\,2390 \hst\ fields)
are indicated by open symbols, elliptical galaxies are denoted by filled
symbols.
Both the distant clusters and the local sample are fitted only within the
region shown by horizontal and/or vertical dotted lines in the plots,
which represent the selection boundaries for the combined A\,2218 and
A\,2390 data. We compare our cluster sample with the local Coma sample by 
\cite{Joerg99} and \cite{JFK95b}. In order to match the selection boundaries
for the distant sample, the Coma galaxies were restricted to $M_{r}<-20.42$ and
log~$\sigma>2.02$.
Under assumption of a non changing slope, we look at the residuals from the
local relation and derive a mean evolution for the distant
galaxies\footnote{For a comparison of the slope of the local and
distant galaxies, we use the bisector method, which is a combination of two
least-square fits with the dependent and independent variables interchanged.
The errors on the bisector fits were evaluated through a bootstrap resampling
of the data 100 times.}.

%
\begin{figure}
\centerline{\psfig{figure=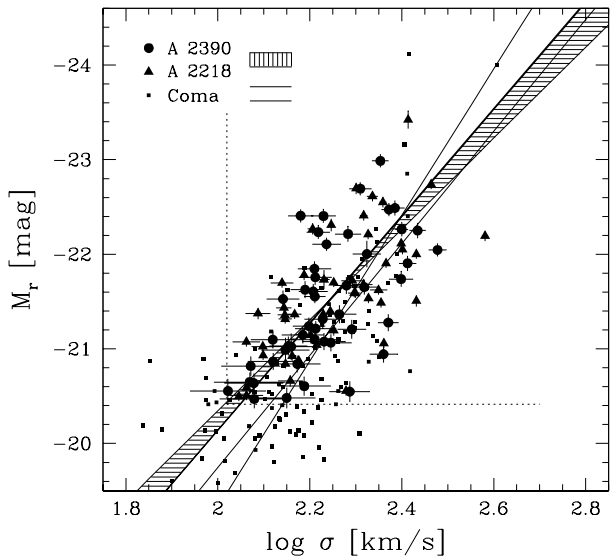,width=80mm}}
\caption{Faber--Jackson relation in Gunn $r$ for 96 early-type galaxies in
A\,2390 and A\,2218, compared to the local Coma sample of J99. The solid
lines show the $\pm1~\sigma$ errors of the 100 iteration bootstrap bisector
fits to the local FJR within selection boundaries,
the hashed area indicates the bisector fits to the distant
sample (within $\pm1~\sigma$).}
\label{FJR}
\end{figure}
%

\subsection{Local comparison sample}

The Coma cluster at $z=0.024$ is one of the best studied local rich clusters.
Therefore, it provides a reliable local reference when addressing evolutionary
questions. \cite{Joerg99} and \cite{JFK95b} (hereafter collectively J99)
performed a detailed study of a large number of early-type Coma galaxies in
Gunn $r$-band. The combined sample comprises 115 early-type galaxies, divided
into subclasses of 35 E, 55 S0 and 25 intermediate types (E/S0). Absolute
magnitudes cover a range down to $M_{r}<-20.02$ with a completeness level
of 93\,\%. In oder to match the local J99 sample, the parameters of the distant
clusters were aperture corrected (see \S\ref{specredu}). A number of recent
works in the literature have analysed the Fundamental Plane (FP) at $z>0.1$ in
the Gunn $r$ and compared it to the Coma cluster
(e.g., \citeauthor{JFHD99} \citeyear{JFHD99}; Z01).
In the intermediate redshift range, observations of early-type galaxies are
preferably in the $R$ or $I$ filters. At $z=0.2$, the observed $I$ and
$I_{814}$ passbands are very close to rest-frame Gunn $r$. Therefore, the
advantages of using the Gunn $r$-band instead of the bluer Johnson $V$ or
$B$ bands are the smaller k-corrections and the lower galactic extinction
corrections.

\subsection{Faber--Jackson relation}

A first test of the formation and evolution of elliptical and lenticular
galaxies allows the scaling relation between galaxy luminosity and velocity
dispersion, the so-called Faber--Jackson relation
(\citeauthor{FJ76} \citeyear{FJ76}). In the $B$-band, the luminosity of an
object is proportional to the random motion of the stars as
$L\propto \sigma^{4}$.

The Faber--Jackson relation (FJR) in the Gunn $r$-band
for 96 distant early-type galaxies in the clusters A\,2218 and A\,2390,
compared to the local Coma sample of J99 is shown in Fig.~\ref{FJR}.
For the FJR scaling relation, the whole cluster data set can be considered as
their magnitudes were measured with SExtractor on the ground-based images.
The distant galaxies have velocity dispersions down to 105~\kms.
A bootstrap bisector fit to the restricted Coma reference sample yields:
\begin{equation}
M_{r}=-6.82\,{\rm log}\,\sigma-5.90.
\end{equation}
The observed $\pm1\sigma$ scatter is $\sigma_{r}=0.57^{m}$. The $\pm1~\sigma$
scatter of the bisector fits to the local FJR is indicated as solid lines in
Fig.~\ref{FJR}. Fits to the distant sample (within $\pm1~\sigma$) are shown
as the hashed area. Because of the younger mean ages of the stellar
populations, the distant clusters are on average brighter than
their local counterparts for a given velocity dispersion.
If we assume a formation redshift of $z=2$ for the cluster galaxies,
the predictions of single-burst passive evolution models
by BC96 (Salpeter IMF with $x=1.35$, mass range $0.10<M_{\sun}<125.0$ and
burst duration of 1~Gyr) suggest an increase of their Gunn $r$ brightness by
0.21$\pm0.05$~mag at $z=0.2$. 
Assuming that the local slope holds valid for our distant galaxies, we analyse
the mean residuals from the local FJR, i.e. the difference
in luminosity for each galaxy from the local FJR fit.
The $\pm1\sigma$ scatter of these offsets for the
distant galaxies are $\sigma_{r}=0.61^{m}$.

Table~\ref{fjrtab} lists the derived mean and median luminosity evolution of
our early-type cluster galaxies for various samples in the FJR.
The final total error in the FJR evolution results from a linear sum
of the total error in absolute magnitude (see~\S\ref{gbima}) and the error 
as introduced by the velocity dispersions as
$\delta \sigma_{M}=a\,{\rm log}\,\delta\sigma_{{\rm err}}$, where
$\delta\sigma_{{\rm err}}$ denotes the mean error in the velocity dispersion
(see Appendix~\S\ref{clusprop}) and $a$ the bisector slope of the
A\,2218 and A\,2390 sub-samples. This error in the mean luminosity evolution
is given as the $\pm$1$\sigma$ uncertainty in Table~\ref{fjrtab}.
For the total sample of 96 galaxies we detect in the
FJR a luminosity evolution of $0.32\pm0.22^{m}$. Similar offsets are deduced
for each cluster separately. The early-type galaxies in A\,2218 show an
evolution of $0.31\pm0.15^{m}$, whereas the E+S0 galaxies in A\,2390 are
brighter by $0.32\pm0.29^{m}$. Both results agree with the BC96 model
predictions. Dividing our sample with respect to luminosity at $M_{r}=-21.49$
in two groups of equal number, the difference in mean luminosity between
lower-luminous and higher-luminous galaxies is negligible. Both sub-samples
have a similar slope with an insignificant slope change ($\Delta a=0.2\pm0.5$).

\subsubsection{Mass dependence}

As the distant galaxies cover a broad range in velocity dispersions
(log\,$2.02\leq\sigma\leq2.58$) down to 105~\kms, it is also possible to
explore any evolution in the slope of the FJR. The Coma sample has a steeper
slope than that of the distant clusters, with a slope difference of 
$\Delta a=1.5\pm0.4$ (see Fig.~\ref{FJR}). This gives some evidence for a
difference between massive and less-massive E+S0 galaxies.
Subdividing the total sample with respect to their velocity dispersion
at $\sigma<170$~\kms (log\,$(\sigma)=2.231$) into two sub-samples of equal
number, the lower-mass objects (log\,$(\sigma)<2.231$) show on average a
larger evolution with $\Delta\,\overline{M}_{r}=0.62\pm0.34^{m}$. More
massive galaxies (log\,$(\sigma)>2.231$) are on average brighter by
$\Delta\,\overline{M}_{r}=0.02\pm0.16^{m}$. Table~\ref{fjrtab} gives
a comparison between the two groups of early-type galaxies.
The galaxies with lower and higher mass feature different mean velocity
dispersions (log\,$(\sigma)<2.15$, and 2.34, respectively). For this reason,
a two-dimensional Kolmogorov-Smirnov (KS) test is not the appropriate
statistical method for a comparison. Instead, we perform bootstrap-bisector
fits to the sub-samples which results in different slopes for the low and
high mass galaxies. The slope difference between the low-mass galaxies and
the high-mass galaxies is $\Delta a=2.5\pm1.5$, which is significant but
depends on the defined selection boundaries for the local Coma reference.
However, there remains the trend that less-massive galaxies feature a larger
luminosity offset which could be a possible hint for a faster evolution of the
low-mass galaxies compared to the more massive counterparts.

\subsubsection{Radial dependence}

In order to investigate any possible dependence on clustercentric radius 
within our large data set, we have measured the distance of each galaxy from
the brightest cluster galaxy and subdivided the cluster sample into two radial
bins of equal size (at $R=458.4$~kpc), so that they comprise equal numbers of
galaxies. The average projected radius for galaxies in the core region is
244 kpc compared to 724 kpc for galaxies in the outer bin.
Based on the relation between the virial mass of a cluster and its 
projected velocity dispersion, the virial radius can be derived
as $R_{v}\,{\rm [Mpc]}\sim 0.0035(1+z)^{-1.5}\sigma_{p}\,h^{-1}$
(\citeauthor{GGMMB98} \citeyear{GGMMB98}),
with the projected velocity dispersion $\sigma_{p}$ given in \kms.
For A\,2218, we adopt $\sigma_{p}=1370^{+160}_{-120}$\,\kms 
(\citeauthor{LPS92} \citeyear{LPS92}) and
for A\,2390 $R_{v}=3.156\,h^{-1}_{100}$~Mpc
(\citeauthor{Car:96} \citeyear{Car:96}). For our assumed cosmology,
this yields to virial radii for the clusters A\,2218 and A\,2390 
of $R_{v}=3.765$ Mpc (1268$''$) and $R_{v}=2.209$ Mpc (605$''$),
respectively.

The mean and median evolution of the E+S0 galaxies for the central and the
outer region are shown in Table~\ref{fjrtab}. 
As the galaxies in the core and outer region feature similar mean velocity
dispersions (log\,$(\sigma)=2.26$, and 2.23, respectively), the
two-dimensional KS test provides a good measure of consistency.
Applying the KS test results in a probability of $p=0.46$ that the two
distributions are similar. The slope for the galaxies in the core
region is similar to the slope derived for the galaxies in the outskirts.
Looking at the residuals from the local FJR, here is a slight trend with
clustercentric radius, but not significantly (within $\pm$1$\sigma$).
On average, galaxies in the outskirts of the cluster
indicate a larger luminosity offset of $0.35\pm0.20^{m}$ 
than in the core region $0.26\pm0.20^{m}$. 
However, the offset has only a low significance and its not sure
whether this gradient is real or maybe just a projection effect. Our galaxies
cover a radial range out to $(R/R_{v})\sim0.5$ (1.1~Mpc), which is in case of
A\,2390 roughly the beginning of the transition region that is far from
the cluster centre but still within the virial radius of the cluster. Therefore,
still many processes are simultaneously at work which could account for
this slight observed trend. Futhermore, our sample
does not reach the outer districts of the clusters such as the complete
transition or periphery zone, where a possible radial gradient with luminosity
would be more apparent. We will readdress this issue in our next paper with
respect to the line strength measurements for our galaxies.

{\scriptsize
\begin{table}
\centering
\begin{minipage}{0.5\textwidth}
\caption{Evolution of the Faber--Jackson relation in Gunn $r$
derived for various samples. $N$ is the number of galaxies and
$\Delta\,\overline{M}_{r}$ indicates the mean luminosity evolution [in mag].
The fourth column denotes the $\pm$1$\sigma$ deviation in the mean luminosity
evolution and the last column $\Delta\,\langle M_{r}\rangle$ gives the
median evolution [in mag].}\label{fjrtab}
\begin{tabular}{lcccc} 
\hline\hline
\noalign{\smallskip}
Sample & $N$ & $\Delta\,\overline{M}_{r}$  &  $\sigma$ &
$\Delta\,\langle M_{r}\rangle$\cr 
\noalign{\smallskip}
\hline
\noalign{\smallskip}
A\,2218          & 48 & 0.31 & 0.15 & 0.29\cr
A\,2390          & 48 & 0.32 & 0.29 & 0.35\cr
A\,2218+A\,2390  & 96 & 0.32 & 0.22 & 0.35\cr
low mass$^{a}$   & 48 & 0.62 & 0.34 & 0.63\cr
high mass        & 48 & 0.02 & 0.16 & 0.01\cr 
in$^{b}$         & 48 & 0.26 & 0.20 & 0.24\cr
out              & 48 & 0.35 & 0.20 & 0.40\cr
\noalign{\smallskip}
\noalign{\hrule}
\end{tabular}
\begin{flushleft}
$^{a}$ less-massive: log\,$(\sigma)<2.231$, more-massive: log\,$(\sigma)>2.231$.\\
$^{b}$ in: only core sample $R<458.4$~kpc,\\ out: only outer
region sample $R>458.4$~kpc.\\
\end{flushleft}
\end{minipage}
\end{table}
}

%
\begin{figure}
\centerline{\psfig{figure=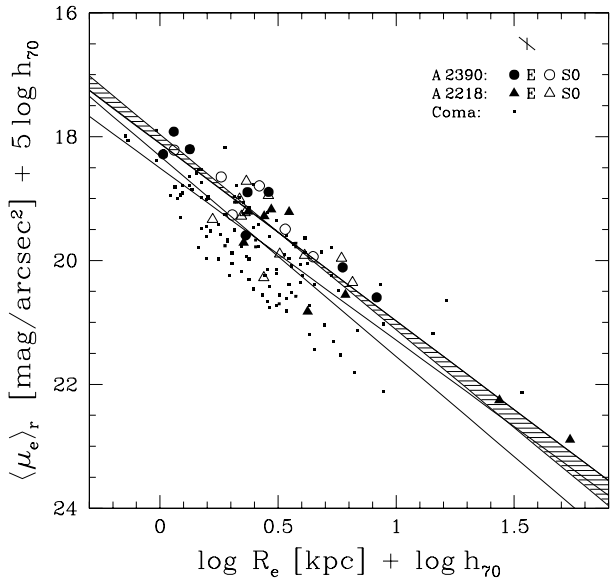,width=80mm}}
\caption{Kormendy relation in the Gunn $r$-band for A\,2390
and A\,2218, compared to the Coma sample of J99. The area bounded by solid
lines indicates the $\pm1~\sigma$ errors of the bisector fits to the local
KR relation. The hashed region shows the bisector fits of the distant sample
within $\pm1~\sigma$. A typical error bar (mean error) is indicated in the
upper right corner (see text for details).}
\label{KRclus}
\end{figure}

\subsection{Kormendy relation}

Another useful probe to study the evolution is the Kormendy relation
(\citeauthor{Korme77} \citeyear{Korme77}), which is the projection of the FP
on the photometric plane. As this relation does not comprise galaxy kinematics,
it can be studied to fainter magnitudes and therefore higher redshifts. However,
the Kormendy relation samples have larger scatter than in the FP and are
more affected by selection biases (\citeauthor{ZSBBGS99} \citeyear{ZSBBGS99}).
Nevertheless, results on the Kormendy relation should complement and endorse
findings obtained with the Fundamental Plane. The Kormendy relation (KR) in
Gunn $r$ for the galaxies in A\,2390 and A\,2218 is shown in Fig.~\ref{KRclus}.
The structural parameters of effective radius $R_{e}$ and mean surface
brightness within $R_{e}$, $\langle\mu_{\rm e}\rangle$, are based on curve of
growth fits and derived using a combination of disk and bulge surface
brightness models. In the upper right corner we show an average
error which enters the KR in the coupled form
${\rm log}\,(R_{{\rm e}})-0.328\,\langle\mu_{\rm e}\rangle=0.08$, assuming a
typical uncertainty of $\Delta\,\langle\mu_{\rm e}\rangle=0.05$ and
$\Delta\,{\rm log}\,(R_{{\rm e}})=0.1$.
In a similar manner as for the FJR, we compute the luminosity evolution for the
distant galaxies in the KR.
Within our selection boundaries, a bisector fit to the
Coma sample gives:
\begin{equation}
\langle\mu_{\rm e}\rangle=3.008\,{\rm log}\,R_{{\rm e}}+18.411
\end{equation}
with an observed scatter of $\sigma_{{\rm KR}}=0.48^{m}$ in
$\langle\mu_{\rm e}\rangle$. 
The $\pm1~\sigma$ scatter of the bisector fits to the local KR 
is shown as solid lines in Fig.~\ref{KRclus}, whereas the bisector fits 
within $\pm1~\sigma$ to the distant sample are indicated as the hashed area.
Again, we assume a constant slope with look--back time for determining the
offsets for the distant sample from the local KR.
The scatter of these offsets is $\sigma_{{\rm KR}}=0.39^{m}$.
Comparing our galaxies within their observed magnitude range  
($17.92\leq \langle\mu_{\rm e}\rangle \leq22.89$)
with the restricted Coma KR, we find a larger luminosity evolution
as with the FJR. In particular, in the rest-frame Gunn $r$ the 34 cluster
galaxies are on average brighter by $0.39\pm0.27^{m}$
(median value 0.46), compared to Coma. Similar results are obtained
using the de Vaucouleurs structural parameters in the KR. The E+S0 galaxies of
both clusters show an offset of $0.38\pm0.27^{m}$, with a median of 0.42.
Note that, if $r^{1/4}$-law structural parameters are used to construct the
KR, \textit{individual} objects are differently distributed along the KR,
whereas the derived \textit{average} offsets from the local KR give similar
results.

\subsection{Fundamental Plane}\label{FP}

The family of early--type galaxies form a homogenous group regarding to
several of their properties.
In a three dimensional parameter space, defined by three observables,
the effective radius $R_{{\rm e}}$, effective surface brightness within
$R_{{\rm e}}$, $\langle\mu_{{\rm e}}\rangle$ and velocity dispersion $\sigma$,
the Fundamental Plane (FP) establishes a tight correlation
(\citeauthor{DD87} \citeyear{DD87}; \citeauthor{Dre:87} \citeyear{Dre:87})
in the following form:
\begin{equation}
{\rm log}\left( \frac{R_e}{{\rm kpc}}\right)\,=\,\alpha\ {\rm log}
\left(\frac{\sigma}{{\rm km~s^{-1}}}\right)\,+\,\beta \left(\frac{\langle\mu_{{\rm e}}\rangle}{{\rm mag~arcsec^{-2}}}\right)\,+\,\gamma.
\end{equation}
This empirical relationship relates galaxy structure ($R_{{\rm e}}$ and
$\langle\mu_{{\rm e}}\rangle$) to kinematics ($\sigma$).

In general, FP studies presume that E+S0 galaxies are a homologous group,
i.e., that they exhibit a similar structure. Under this assumption the total
galaxy mass (including dark matter) is proportional to its virial mass
$\sigma^{2}\,R_{e}\,G^{-1}$ (\citeauthor{Treu01b} \citeyear{Treu01b}).
Elliptical galaxies do not fill the FP plane entirely, but rather are
restricted to a certain band within it (\citeauthor{Guz:93} \citeyear{Guz:93}).
Different factors account for the precise form of the FP, known as the lack of
exact homology of early-type galaxies, e.g., visible in differences in the
luminosity profiles or in the dynamical structure
(\citeauthor{Cao93} \citeyear{Cao93}; \citeauthor{Gra:96} \citeyear{Gra:96}).
Thanks to the small scatter of the local FP of $\sim$0.1~dex, the formation and
evolution of E+S0 galaxies can be constraint with good precision
(e.g., \citeauthor{BBF92} \citeyear{BBF92};
\citeauthor{PDC98a} \citeyear{PDC98a}).

\begin{figure*}
  \centering
  \includegraphics[width=0.9\textwidth,angle=0]{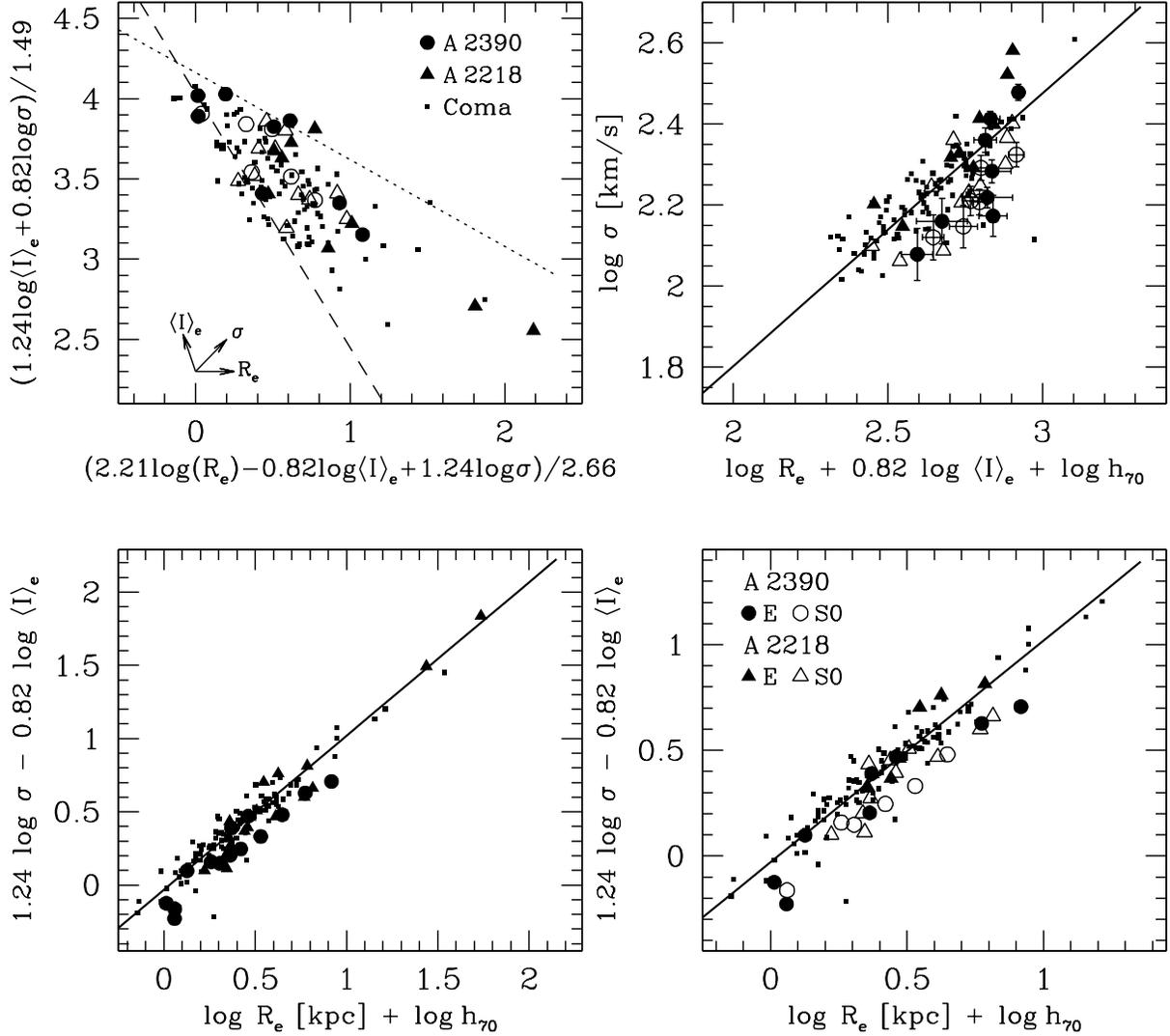}
  \caption{Fundamental Plane for A\,2390 at $z=0.23$ (circles) and A\,2218
  at $z=0.18$ (triangles) in rest-frame Gunn $r$-band, compared to the Coma
  galaxies of J99 (small squares). Filled symbols denote ellipticals, open
  symbols S0 galaxies and Sa bulges.
  {\it Upper panel, left:} Face on FP. The dotted line indicates the so-called
  `exclusion zone' for nearby galaxies (Bender et al. 1992), the dashed line
  the luminosity limit for the completeness of the Coma sample
  $M_{r\,{\rm T}}=-20.02^{{\rm m}}$. {\it Upper panel, right:} FP edge-on,
  along short axis. The solid line represents the bisector fit for the local
  Coma sample. {\it Lower panel:} Edge-on FP. On the right
  side a zoom of the edge-on FP with a separation into different morphologies
  is shown.}
  \label{FPr}
\end{figure*}

The Fundamental Plane in rest-frame Gunn $r$ for the distant clusters
A\,2218 ($z=0.175$) and A\,2390 ($z=0.228$) is illustrated in Fig.~\ref{FPr}. 
The figure also shows the FP for the local Coma samples of J99 ($z=0.024$).
All mean surface brightness magnitudes have been corrected for the dimming due
to the expansion of the Universe. Errors are provided in the short edge-on
FP projection (upper right panel of Fig.~\ref{FPr}), with kinematic and
photometric properties on separate axes.
Following results were derived. Firstly, the FP for both individual galaxy
populations as well as for the combined sample show a well-defined, tight
relation with a small scatter. Both intermediate redshift clusters show a
similar behaviour within and along all projections of the FP. There is no
evidence for an increasing scatter with redshift.
\cite{JFK96} found for a sample of 10 local clusters an rms scatter
of $\sigma=0.084$ in log~$R_{e}$. Our distant cluster galaxies have an rms
scatter of $\sigma=0.113$, which is not significant higher than the local
value. Secondly, only a moderate evolution of the FP and hence the stellar
populations of the galaxies with redshift is seen. Assuming there is
no evolution in the structure of the galaxies, i.e., at a fixed $R_{e}$ and
$\sigma$, the average brightening of the cluster galaxies can be determined.

The results for the FP in rest-frame Gunn $r$ are presented in
Table~\ref{fptab}. We show the mean and median zero-point offsets from the local
Coma FP, their 1$\sigma$ scatter and the derived luminosity
evolution. For the combined sample of 34 early-type cluster galaxies we deduce
an ZP offset of $\Delta \gamma=0.10\pm0.06$ compared to the local Coma
reference, which corresponds to a modest luminosity evolution of
$\Delta\,\mu_{\rm e}=0.31\pm0.18$~mag.
Our results are consistent with the picture of simple passive evolution models
(e.g., BC96). Assuming a formation redshift of $z_{\rm f}=2$ for all stars, for
example, these models predict a brightening by $\Delta m_{r}\approx 0.20$~mag.
For the individual clusters, different results are obtained.
A zero-point offset of $0.16\pm 0.06$ for A\,2390 alone is found, meaning that
the early-type galaxies in A\,2390 are on average more luminous by
$\Delta\,\mu_{\rm e}=0.49\pm0.18$~mag than the local Coma sample of J99.
For the E+S0 in A\,2218 we detect an offset of $\Delta \gamma=0.03\pm0.06$,
corresponding to a brightening of the stellar populations by
$\Delta\,\mu_{\rm e}=0.09\pm0.18$~mag. This different result for the clusters
may be caused by a combination of two effects: ({\it i}) possible effect due to
cosmic variance and ({\it ii}) sample selection. For example, \cite{JFHD99} have
analysed two different clusters (A\,665 and A\,2218) both at a redshift of
$z=0.18$ and found a slight different evolution of $\Delta\,\mu_{\rm e}\sim0.15$~mag. This
may give evidence that not all rich clusters have the same FP zero-point offset
and that cosmic variance must be accounted for.
Another reason may arise from our sample selection. In case of A\,2390
particularly fainter galaxies have been selected in order to gain additional
insights on the low-mass end of the FP at $z\sim0.2$. For this reason, the
A\,2390 FP sub-sample comprises more low-luminous galaxies than that of A\,2218,
which results on average in a stronger luminosity evolution for the A\,2390
cluster.  Table~\ref{fptab} gives a comparison between the two
sub-samples of less and more-massive galaxies and between faint and bright
galaxies. If we divide the total sample regarding to velocity dispersion
at $\sim$192~\kms (corresponding to log\,$(\sigma)<2.283$) into two
sub-samples of different mass, we derive a stronger evolution by 
$\Delta\,\mu_{\rm e}=0.47\pm0.24$~mag for the low-mass
galaxies with respect to their more massive counterparts
$\Delta\,\mu_{\rm e}=0.03\pm0.21$. Similar results are
found when we make a cut in luminosity at $M_{r}=-21.493$,
although with a larger scatter. 
Looking at the thickness of the FP with respect to luminosity, we do not find an
increasing scatter for the fainter galaxies within our sample. However, our
sub-sample is too small to test the scatter of the FP in greater detail. 
\cite{JFK96}, for example, divided their local sample with respect to luminosity
into faint ($M_{r}>-23.16^{{\rm m}}$) and bright galaxies and reported on a
larger rms scatter for the lower-luminous galaxies of 0.086. They argued that
the larger scatter may be a result of the presence of disks or larger
variations in the stellar populations of lower luminosity galaxies.
The difference between the lower-luminous and higher luminous galaxies in our
sample is also visible in the size distribution of the two samples
(Fig.~\ref{cluscomp}) and in the Kormendy relation, with A\,2390 containing
more galaxies with small sizes and three even with ${\rm log}\,R_{e}\leq0.05$.

{\scriptsize
\begin{table}
\centering
\begin{minipage}{0.5\textwidth}
\caption{Evolution of the FP in Gunn $r$ as derived for various samples.
$N$ shows the number of galaxies and $\Delta\,\overline{\gamma}$ indicates the
mean FP zero-point offset. In the fourth and fifth column,
the median FP zero-point evolution $\Delta\,\langle \gamma\rangle$ and the 
median evolution in the FP $\Delta\,\langle \mu_{\rm e}\rangle$ [in mag] are
listed. The last column gives the $\pm$1$\sigma$ scatter
of the mean offsets.}\label{fptab}
\begin{tabular}{lccccc} 
\hline\hline
\noalign{\smallskip}
Sample & $N$ & $\Delta\,\overline{\gamma}$  & 
$\Delta\,\langle \gamma\rangle$ & $\Delta\,\langle \mu_{\rm e}\rangle$ &
$\sigma$ \cr
\noalign{\smallskip}
\hline
\noalign{\smallskip}
A\,2218          & 20 & 0.032 & 0.026 & 0.079 &  0.108 \cr
A\,2390          & 14 & 0.140 & 0.158 & 0.482 &  0.088 \cr
A\,2218+A\,2390  & 34 & 0.076 & 0.100 & 0.305 &  0.113 \cr
\hline
E$^{a}$          & 17 & 0.032 & 0.008 & 0.024 &  0.122 \cr
S0$^{a}$         & 17 & 0.120 & 0.144 & 0.439 &  0.085 \cr
E (pure $r^{1/4}$)$^{b}$ & 17 & 0.032 & -0.010 &  -0.030 & 0.125 \cr
S0 (pure $r^{1/4}$)$^{b}$ & 17 & 0.115 & 0.134 & 0.409 & 0.079 \cr
\hline
low lum.$^{c}$   & 17 & 0.085 & 0.100 & 0.305 & 0.114 \cr
E                &  8 & 0.070 & 0.074 & 0.226 & 0.127 \cr
S0               &  9 & 0.098 & 0.122 & 0.372 & 0.107 \cr
high lum.        & 17 & 0.068 & 0.093 & 0.284 & 0.114 \cr
E                &  9 &-0.001 &-0.020 &-0.061 & 0.114 \cr
S0               &  8 & 0.145 & 0.168 & 0.512 & 0.045 \cr
\hline
low mass$^{d}$   & 17 & 0.129 & 0.153 & 0.466 & 0.106 \cr
E                &  6 & 0.114 & 0.158 & 0.482 & 0.150 \cr
S0               & 11 & 0.138 & 0.153 & 0.466 & 0.080 \cr
high mass        & 17 & 0.023 & 0.010 & 0.030 & 0.095 \cr
E                & 11 &-0.012 &-0.015 &-0.046 & 0.081 \cr
S0               &  6 & 0.087 & 0.100 & 0.305 & 0.091 \cr
\noalign{\smallskip}
\noalign{\hrule}
\end{tabular}
\begin{flushleft}
$^{a}$ based on $r^{1/4}$+exp. disk surface brightness profiles.\\ 
$^{b}$ $r^{1/4}$ surface brightness profile only.\\
$^{c}$ lower-luminous: $M_{r}>-21.493$, higher-luminous: $M_{r}<-21.493$.\\
$^{d}$ less-massive: log\,$(\sigma)<2.283$, more-massive: log\,$(\sigma)>2.283$.\\
\end{flushleft}
\end{minipage}
\end{table}
}

\begin{figure*}
\vspace*{0.3cm}
\centerline{\psfig{figure=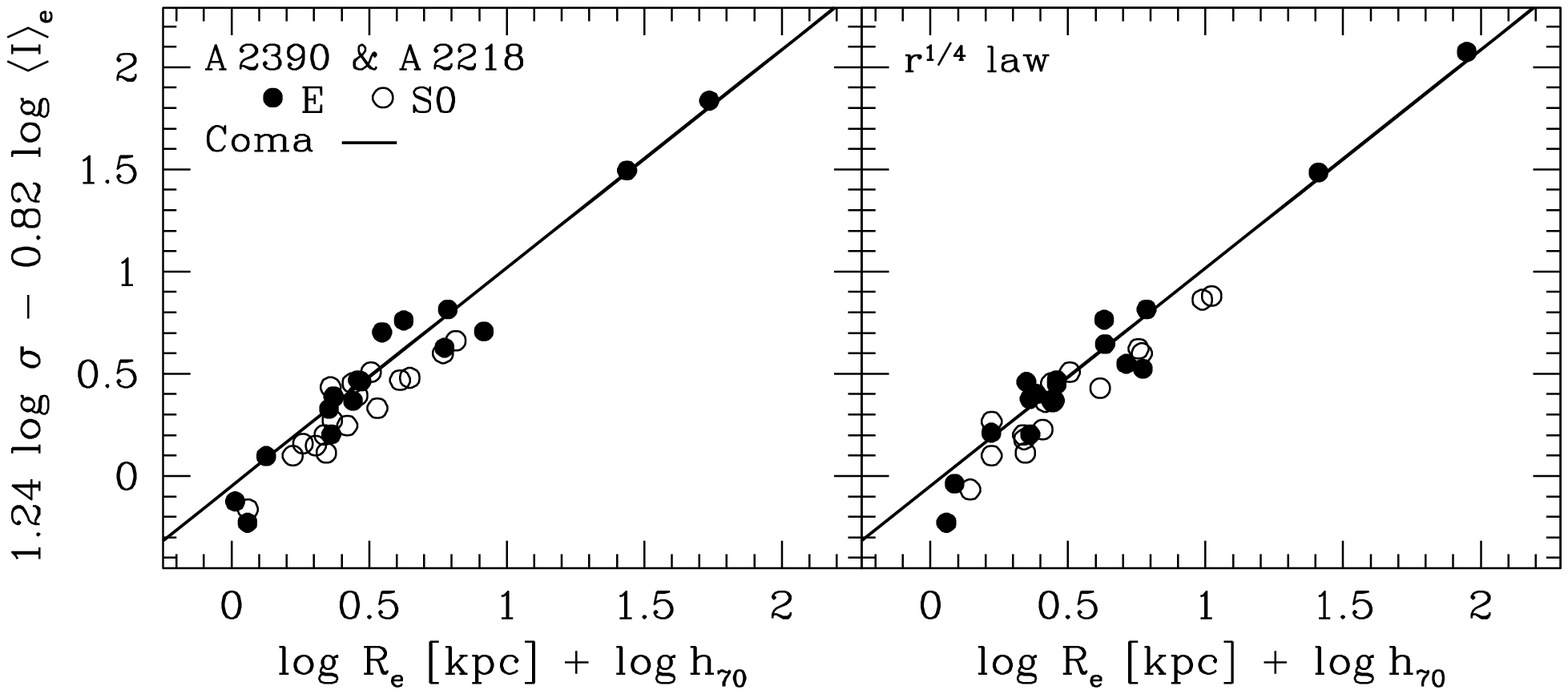,width=140mm}}
\caption{Edge-on view of the Fundamental Plane for A\,2390 and A\,2218 in
rest-frame Gunn~$r$. The distant FP is compared to the FP of the local Coma
sample of J99, indicated by the 100 iteration bootstrap bisector fit
(thick solid line). {\it Left:}\/ FP constructed using
a combination of an $r^{1/4}$ and exponential disk profile. {\it Right:}\/
Similar to the left plot except that $r^{1/4}$-law parameters were
used.}
\label{FPedge}
\end{figure*}

\subsubsection{Elliptical versus lenticular galaxies}

Local investigations based on the Fundamental Plane have not revealed
significant differences in the zero-point, slope and/or scatter between
elliptical and lenticular galaxies (\citeauthor{BBF92} \citeyear{BBF92},
\citeyear{BBF93}; \citeauthor{SBD93} \citeyear{SBD93};
\citeauthor{JFK96} \citeyear{JFK96}). Moreover, they behave very similar as
one single group of galaxies with respect to their $M/L$ ratios
and within the FP. Going to higher redshifts, differences could be more
significant if recent star formation activity plays a role for S0 galaxies.
However, previous studies of samples at higher
redshift did not find any differences between elliptical and S0 galaxies
(\citeauthor{DF96} \citeyear{DF96}; \citeauthor{KDFIF97} \citeyear{KDFIF97};
\citeauthor{DFKI98} \citeyear{DFKI98}).
For example, within the large sample of 30 early-type
galaxies in CL\,1358+62 at $z=0.33$ of \cite{KIDF00}, the 11 ellipticals
displayed identical zero-points as the 13 (non--E$+$A) S0 galaxies with no hint
for an offset between these groups at all. Moreover, the difference in the
slope of $\sim$14\,\% detected between the S0 and elliptical galaxies is
not significant.

For the case of S0 galaxies, two main questions are still a matter of debate.
How many are a priori S0 galaxies? Which and how many are the result of galaxy
transformations and account for the dominant S0 population in rich clusters
today? For scenarios which suggest a transformation of star-forming spiral
galaxies into passive S0 systems (\citeauthor{DOCSE97} \citeyear{DOCSE97};
\citeauthor{KSNOB01} \citeyear{KSNOB01}), these objects can only be classified
as lenticular after their morphology has been changed ($\sim$5 Gyr). However,
after this relatively long time-scale, their star formation (SF) could have
already been ceased which would make these galaxies hard to detect.
E$+$A galaxies could represent galaxies in an intermediate stadium of such a
transformation where harassment just has stopped but SF is still present due
the result of a starburst which ended within the last 1.5--2 Gyr
(\citeauthor{BAECSS96} \citeyear{BAECSS96}). Recently, \cite{YZZea04}
performed a study of E$+$A galaxies in the local Universe and argued that these
objects only shown an offset in their surface brightnesses but not in their
total magnitudes. Such galaxies would be invisible in the FJR but could
be revealed within the FP.

Figure~\ref{FPedge} illustrates the edge-on view of the FP for 34 E+S0 galaxies
of the clusters A\,2218 and A\,2390 in rest-frame Gunn $r$-band. The FP
constructed for the distant clusters at intermediate redshift is compared to
the FP for the local Coma sample of J99, indicated by the bisector fit.
Our combined sample represents a comparable data set to the study
by \cite{KIDF00} and hence one of the most extensive
investigations of early-type galaxies in clusters at $z\sim 0.2$.
The Fig.~\ref{FPedge} has been splitted in order to visualise the small
variations if different luminosity profile fits are used for deriving structural
parameters. The left figure displays the FP constructed using a combination of
an $r^{1/4}$-law$+$exponential disk profile, the right figure is based on pure
$r^{1/4}$-law fits. The variations in the structural parameters only affect the
galaxies to move along the edge-on projection of the FP plane, thereby
maintaining the tightness of the plane. As shown by \cite{GC97}, both the
$r^{1/4}$ FP and $r^{1/n}$ FP have the same scatter and slope within their
errors. A second comparison is given by the FP parameter
$R_{{\rm e}}\,I_{e}^{0.8}$ in Table~\ref{hstpar} and
Table~\ref{dVhstpar} (column 9). In general, the agreement
is good for both elliptical and S0 galaxies.

We will now consider in more detail possible differences between elliptical
and lenticular galaxies. As a local reference we adopt the Coma FP coefficients
of J99 in the Gunn $r$-band. The choice of
the fitting technique, the selection criteria and the measurement errors which
are correlated can lead to systematic uncertainties in the FP coefficients in
the order of $\pm$0.1~dex (\citeauthor{JFK96} \citeyear{JFK96};
\citeauthor{KIDF00} \citeyear{KIDF00}).
A morphological analysis of the \hst\ images revealed that our A\,2390
sub-sample splits nearly equally into elliptical (8) and lenticular (S0)
galaxies (6). Both ellipticals and lenticular galaxies are uniformly distributed
along the surface of the FP plane. An edge-on projection can therefore be taken
for a robust comparison of their stellar populations. Table~\ref{fptab} lists
the derived evolution for the E and S0 galaxies. Fixing the slope of the local
Coma reference, the zero-point offset for the S0 galaxies in our sample is:
\begin{equation}
\Delta \gamma_{{\rm S0\, ,\, z=0.2}}=0.14\pm0.06
\end{equation}
This zero-point offset for lenticular galaxies corresponds to an evolution
of $\Delta\,\mu_{\rm e}=0.44\pm0.18$~mag with respect to the local
counterparts. 
On the other hand, the ZP offset for the ellipticals yields:
\begin{equation}
\Delta \gamma_{{\rm E\, ,\, z=0.2}}=0.01\pm0.07
\end{equation}
Restricting our sample to the elliptical galaxies we derive a negligible
zero-point deviation with respect to the local FP, which corresponds to
a insignificant luminosity evolution of
$\Delta\,\mu_{\rm e}=0.02\pm0.21$~mag which is within their
$1\sigma$ scatter.
For the combined sample of 34 E+S0 cluster galaxies we derive a median
offset of:
\begin{equation}
\Delta \gamma_{{\rm E+S0\, ,\, z=0.2}}=0.10\pm0.06,
\end{equation}
which corresponds to an brightening of the stellar populations by
0.31$\pm$0.18~mag.
On average, the lenticular galaxies in our sample show a larger luminosity
evolution than the ellipticals. In both interpretations of Fig.~\ref{FPedge}
the S0 galaxies are predominantly located below the ellipticals and may
indicate a different evolutionary trend between the stellar populations of
elliptical and S0 galaxies.
Dividing our sample with respect to velocity dispersion into a low mass
(log\,$(\sigma)<2.283$)
and a high mass sub-sample (log\,$(\sigma)>2.283$), we find a different
evolution. The lower-mass galaxies are on average with
$0.47\pm0.24^{\rm m}$
more luminous than their more massive counterparts
with $0.03\pm0.21^{\rm m}$ at $z\sim 0.2$. This is also seen for the different
sub-groups of elliptical and S0 galaxies, but with less significance (see 
Table~\ref{fptab}).

In conclusion, we derive for the whole sample of 34 E+S0 galaxies only a
moderate amount of luminosity evolution. From these results and the previous
findings of the FJR and KR, we conclude that at a look-back time of
$\sim$2.8~Gyrs most early-type galaxies of A\,2218 and A\,2390 consist of an
old stellar population with the bulk of the stars formed at a high formation
redshift of about $z_{{\rm f}}\ge 4$.


\subsection{The $M/L$ ratio as diagnostics for stellar populations}\label{mlr}

Under assumption of homology of the early-type galaxies, the existence of the
tight FP implies that the global mass-to-light ($M/L$) ratio of early-type
galaxies is related with the mass $M$ of an object in the following form
(\citeauthor{Fab:87} \citeyear{Fab:87}):
\begin{equation}
M/L_{r}\propto \sigma^{0.49}\, R_{e}^{0.22}\propto M^{0.24}\, R_{e}^{-0.02}
\end{equation}
The small scatter of $\pm$0.10 dex within the FP implies a scatter of only 23\%
in their $M/L$ ratios.

The evolution of the Fundamental Plane can be characterised in terms of its
zero-point. In turn, the zero-point of the FP is related to the mean $M/L$
ratio. Thus, if the FP zero-point for a sample of early-type galaxies
changes as a function of redshift $z$, this implies an evolution of the mean
$M/L$ ratio and hence an evolution in the luminosity-weighted stellar population
properties of the galaxies under consideration.

For the present-day zero-point of the FP we performed a bootstrap bisector fit
to the early-type Coma galaxies in rest-frame Gunn $r$-band. This yield a slope
of $1.048\pm0.038$ and a zero-point of $0.412\pm0.010$.
The early-type galaxies of A\,2390 show a small shift with respect to the Coma
galaxies (see Fig.~\ref{FPr}). 
As the velocity dispersions and effective radii of the galaxies in A\,2390
span similar ranges like their local counterparts (apart from the Coma cD
galaxies with $R_{e}\ga 30$~kpc) and the surface brightnesses are slightly
increased with respect to the local sample, at least some of the shift in the
$M/L$ ratio is due to luminosity evolution.
However, when deriving an exact amount of luminosity evolution, one
has to be careful as additional combined effects also contribute to the
total derived luminosity offset. Some studies found different FP slopes at
intermediate redshift compared to the local slope, but they suffer from low
number statistics (\citeauthor{DF96} \citeyear{DF96};
\citeauthor{KDFIF97} \citeyear{KDFIF97}; \citeauthor{DFKI98} \citeyear{DFKI98}).
However, more recent investigations based on larger samples reported similarly
only a modest change in the FP slope (\citeauthor{KIDF00} \citeyear{KIDF00};
\citeauthor{Treu01b} \citeyear{Treu01b}; Z01).
Thus, the lack of a strong slope change gives evidence against the hypothesis
of the FP being solely an age-mass relation.
The Fundamental Plane is mainly a relation between the $M/L$ ratio and galaxy
mass $M$. Therefore, when comparing small samples with significantly different
ranges in galaxy mass, an offset in the $M/L$ ratio can heavily depend on the
adopted FP slope. The $M/L$ offset will be a combination of three items:
({\it i}) the difference in the slope of the FP, ({\it ii}) the mean
mass range of the sample and ({\it iii}) the `true' offset. By reducing the
differences in the mass distribution, the $M/L$ results become insensitive to
the adopted slope of the FP and any derived evolution for the distant galaxies
is valid for a given range of galaxy masses
(\citeauthor{KIDF00} \citeyear{KIDF00}). For this reason, we have restricted
the local Coma sample to a similar mass function as for the galaxies in
A\,2218 and A\,2390.

%
\begin{figure}
\centerline{\psfig{figure=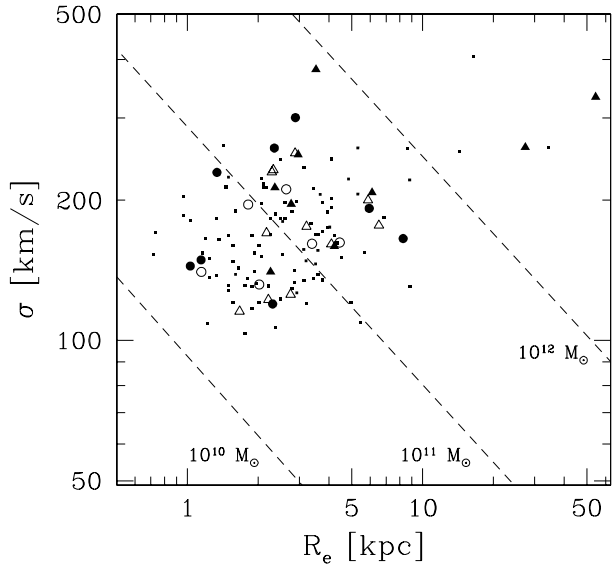,width=80mm}}
\caption{The ${\rm log}\ \sigma-{\rm log}\ R_{e}$ plane for the A\,2390
and A\,2218 early-type galaxies, compared to the Coma galaxies of J99.
Symbol notations as in Fig.~\ref{FPr}. Dashed lines indicate contours of
constant mass $5G^{-1}\,\sigma^{2}\,R_{{\rm e}}=10^{10}$, $10^{11}$ and
$10^{12}\,M_{\sun}$ (Bender et al. 1992). The early-type galaxy masses of
A\,2390 and A\,2218 at $z\sim 0.2$ are similarly distributed as the
early-type Coma sample.}
\label{sigre}
\end{figure}

A comparison of the mass distributions of the distant early-type galaxies of
A\,2390 and A\,2218 with the local Coma galaxies within the
${\rm log}\,\sigma-{\rm log}\,R_{e}$ plane is shown in Fig.~\ref{sigre}.
Regions of constant mass, ranging from $10^{12}\,M_{\sun}$ over
$10^{11}\,M_{\sun}$ down to $10^{10}\,M_{\sun}$, derived with the relation
$M=5\,\sigma^{2}\,R_{{\rm e}}/G$ in units of $M_{\sun}$
(\citeauthor{BBF92} \citeyear{BBF92}) where $G$ is the gravitational constant,
are indicated as dashed lines.
Both samples exhibit similar ranges in mass ($\sigma$) and size ($R_{e}$),
assuring that a possible $M/L$ evolution will not be driven mainly by any
differences between the galaxy mass ranges of the samples.

Adopting a constant slope with redshift, the median zero-point offset
for the combined sample of 34 E+S0 cluster galaxies yielded
$\Delta \gamma_{{\rm E+S0\, ,\, z=0.2}}=0.10\pm0.02$ (where the error
corresponds to the uncertainty as derived via the bootstrap method).
Alternatively using a variable slope in the FP for the distant
cluster sample, we find negligible changes in the FP parameters $\alpha$ and
$\beta$ compared to the locally defined parameters by the Coma data.
In particular, a bootstrap bisector fit to all 34 E+S0 cluster gives
a slightly steeper slope of $1.152\pm0.047$ and a zero-point offset of
$0.024\pm0.018$, with a 1$\sigma$ confidence level for a slope change.
This comparison and the results based on the $M/L$ ratios may give some
evidence for a mass-dependent evolution which is stronger for low-mass
galaxies.

%
\begin{figure*}
\centerline{\psfig{figure=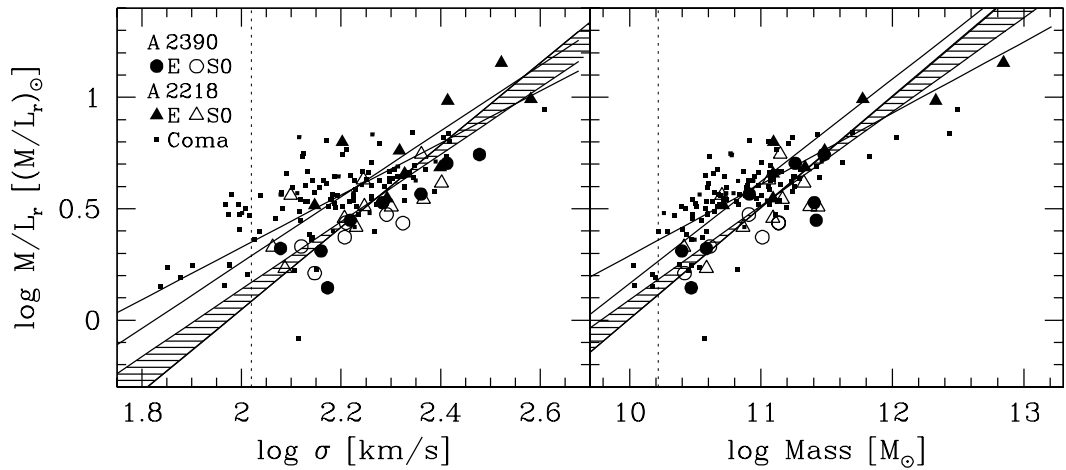,width=140mm}}
\caption{The $M/L$ ratio in Gunn $r$ for A\,2390 and A\,2218
as a function of velocity dispersion $\sigma$ (left) and as a function of
mass in solar units (right). Symbol notations as in Fig.~\ref{FPr}.
Dotted lines show the 100 iteration bootstrap bisector fits to the local Coma
sample of J99 within selection boundaries of the distant samples
(1~$\sigma$ errors). Solid lines are the bisector fits within 1~$\sigma$
errors to the combined distant sample of A\,2390 and A\,2218.}
\label{MLall}
\end{figure*}
%

Fig.~\ref{MLall} shows the observed mass-to-light $M/L$ ratio of the distant
clusters A\,2218 and A\,2390 as a function of velocity dispersion $\sigma$
(left) and versus mass $M$ (right). We limited the Coma sample to
galaxies with log~$\sigma>2.02$ (indicated by the vertical dotted lines in
Fig.~\ref{MLall}) in order to match the area of parameter space covered by the
distant galaxies of A\,2390 and A\,2218.
An analysis based on bootstrap bisector
fits to $M/L=a\,\sigma^{m}$ revealed different $M/L$ slopes for the distant
(solid lines) and the local (dotted lines) samples.
The slope difference between A\,2218 and Coma is
$\Delta m_{\rm A2218}=0.27\pm0.17$ and an
offset in the zero-point of $\Delta a_{\rm A2218}=0.040\pm0.027$. For A\,2390
we detect $\Delta m_{\rm A2390}=0.11\pm0.21$ and
$\Delta a_{\rm A2390}= 0.157\pm0.019$. The combined sample of distant clusters
has a slope difference of $\Delta m_{\rm z=0.2}=0.36\pm0.17$ and
$\Delta a_{\rm z=0.2}=0.036\pm0.024$. We measure with a $2\sigma$ confidence
different slopes for both, the intermediate clusters and the Coma cluster
(Coma slope value is $m=0.59\pm0.15$). However, we detect a
systematic zero-point offset of the
distant $M/L$ relation of $\Delta a_{\rm z=0.2}=0.04\pm0.02$, which is not
agreement with the proposed change due to passive evolution with $z_{\rm f}=2$
($\Delta a=0.12$). Ellipticals and S0 galaxies seem to have different
$M/L$ slopes changes, 
with a steeper slope for the elliptical galaxies. Ellipticals:
$\Delta m_{\rm E}=0.54\pm0.30$ and $\Delta a_{\rm E}=0.041\pm0.037$; S0s:
$\Delta m_{\rm S0}=0.06\pm0.23$ and $\Delta a_{\rm S0}=0.141\pm0.026$.

The $M/L$ ratios as a function of mass $M$ are shown in the right plot of
Fig.~\ref{MLall}. In a similar manner as for the velocity dispersions,
the two samples are analysed in terms of bootstrap bisector fits.
For the Coma cluster alone we find is $M/L\propto M^{0.59\pm0.07}$. 
With respect to Coma, the slope difference for A\,2218 is 
$\Delta m_{\rm A2218}=0.025\pm0.060$ with a zero-point offset of
$\Delta a_{\rm A2218}=0.048\pm0.029$. For A\,2390 a similar slope change
is found $\Delta m_{\rm A2390}=0.057\pm0.063$,
but a larger offset of $\Delta a_{\rm A2390}=0.165\pm0.028$.
The two distant clusters as a whole have $\Delta m_{\rm z=0.2}=0.071\pm0.039$
and $\Delta a_{\rm z=0.2}=0.037\pm0.021$. Dividing the sample into elliptical
and S0 galaxies gives no significant slope changes, for Es:
$\Delta m_{\rm E}=0.049\pm0.056$ and for S0s: $\Delta m_{\rm S0}=0.087\pm0.089$.
However, the S0s exhibit a larger offset in the zero-point of
$\Delta a_{\rm S0}=0.138\pm0.030$, compared to the ellipticals
$\Delta a_{\rm E}=0.050\pm0.034$.

The correction of the so-called `progenitor bias'
(\citeauthor{vD01b} \citeyear{vD01b}) for our sample has insignificant
influence on the results in the FP and on the $M/L$ ratios (at $z=0.2$
the evolution in the $M/L$ ratio is underestimated by
$\Delta {\rm ln}\,\langle M/L_{B}\rangle\approx0.2\times z\approx0.04$).
Therefore, we have neglected it. This effect comes into play at a redshift of
approximately $z\ga0.4$ and has dramatic effect at high redshifts of about
$z\ge0.8$.



\section{Conclusions}

We have investigated the evolution for a total of 96 early-type galaxies in
two massive, $X$-ray luminous clusters of galaxies, Abell~2218 ($z=0.175$) and
Abell~2390 ($z=0.228$). The data of Abell~2218 was already described
in Z01, whereas new kinematics and structural parameters of 48
early-type galaxies in the cluster Abell~2390 are presented. These combined
data set represents one of the largest samples at redshift of $z\sim 0.2$,
ideally suited to study the evolution of early-type cluster galaxies. For the
whole sample we have constructed the Faber-Jackson relation (FJR) and for a
sub-sample of 34 early-type galaxies, we have explored the evolution of the
Fundamental Plane (FP) and Kormendy relation (KR).
Our main results are as following:
\begin{itemize}
\item In the FJR we find a modest luminosity evolution of the
96 early-type galaxies with respect to the local reference.
For the total sample the average offset from the local FJR in the Gunn $r$-band
is $\Delta\,\overline{M}_{r}=0.32\pm0.22^{m}$.
Similar findings are obtained if we restrict the sample to the
individual clusters. A\,~2390 has a mean brightening of
$0.32\pm0.29^{m}$, A\,~2218 shows an evolution of $0.31\pm0.15^{m}$.
Both results agree with the BC96 model predictions for a passive evolution of
the stellar populations assuming a formation redshift of $z_{\rm f}=2$.
\item Splitting the total sample of the FJR with respect to velocity dispersion,
less-massive galaxies ($\sigma<170$~\kms) show a trend for a larger
evolution with $\Delta\,\overline{M}_{r}=0.62\pm0.34^{m}$
compared to their more massive counterparts
$\Delta\,\overline{M}_{r}=0.02\pm0.16^{m}$.
There is little evidence for a small gradient with clustercentric 
radius with slightly brighter luminosities for galaxies in the outer regions.
\item The 34 early-type cluster galaxies obey a tight Fundamental
Plane relation with a small scatter of 0.11~dex at $z\sim0.2$. We find
no evidence for an increasing scatter with redshift. For the combined cluster
samples we detect a mild luminosity evolution with a zero-point offset of
$\Delta \gamma_{{\rm z=0.2}}=0.10\pm0.06$ with respect to the local Coma FP
relation.
The cluster A\,2390 alone indicates a larger evolution of
$\Delta \gamma_{{\rm A\,2390}}=0.16\pm0.06$. However, this is due to a larger
number of low-mass galaxies in the case of A\,2390. For A2218 we derive
$\Delta \gamma_{{\rm A\,2218}}=0.03\pm0.06$.
\item Subdividing our sample in terms of morphology yielded two sub-samples of
elliptical (E) and lenticular (S0) galaxies which are equally in size.
Ellipticals and S0 galaxies are uniformly distributed along the FP
(Fig.~\ref{FPr}), with a similar scatter of 0.1~dex. The sub-samples
show a median zero-point offset from the local reference of
$\Delta \gamma_{{\rm E}}=0.01\pm0.07$ and
$\Delta \gamma_{{\rm S0}}=0.14\pm0.06$, respectively.
We find little evidence for differences between elliptical and
S0 galaxies within our sample. Lenticular galaxies alone induce
on average a larger evolution thereby residing a zone in the FP
preferably below the ellipticals.
\item An analysis of the $M/L$ ratios for the distant galaxies revealed
a steeper slope than that of the local counterparts. This effect is weaker than
that derived by \cite{JFHD99} for a composite of five intermediate redshift
clusters. The sub-samples of ellipticals and S0s appear to have different
$M/L$ slopes, with a trend for a steeper slope for the elliptical galaxies
(Fig.~\ref{MLall}). The $M/L$ ratios as a function of mass $M$ indicate no
differences in the slopes between elliptical and lenticular galaxies.
However, the S0 galaxies feature a larger zero-point offset.
\end{itemize}

Our results are in good agreement with previous observational studies on rich
clusters at intermediate and higher redshifts
(\citeauthor{JFHD99} \citeyear{JFHD99}; \citeauthor{KIDF00} \citeyear{KIDF00};
\citeauthor{WvDKFI04} \citeyear{WvDKFI04}). These investigations suggest that
the mild luminosity evolution derived for their galaxies is consistent with
stellar population models of passive evolution.
Assuming a formation redshift of $z_{\rm f}=2$, the BC96 models predict in the
Gunn $r$-band a brightening by 0.21~mag at $z=0.2$. This is in
good agreement with the findings we derive for the FJR and FP. For the cluster
A\,2390 alone, we detect a larger evolution in the FP and KR. This may be the
result of a combination of both cosmic variance and of a larger number of
low-luminous galaxies. Our results are consistent with the monolithic collapse
scenario and a passive evolution of the stellar populations.

In a forthcoming paper, we will explore the line strengths for our early-type
galaxies in order to analyse the stellar populations of the E+S0 cluster
galaxies with respect to age, metallicity and abundance ratios. Results of the
line indices will be compared to stellar population models and will reveal if
there are different sub-populations between elliptical and lenticular galaxies
with respect to their age or metallicity of a combination of both properties.


\section*{Acknowledgments}

We thank the anonymous referee for constructive criticism that improved the
clarity of this manuscript. We would like thank Prof. K.~J. Fricke for
encouragement and support and are grateful to Drs. A. B\"ohm and A.~W.~Graham
for stimulating discussions and valuable comments.
We thank the Calar Alto staff for efficient observational support.
AF and BLZ acknowledge financial support by the Volkswagen Foundation
(I/76\,520) and the Deutsche Forschungsgemeinschaft (ZI\,663/1-1, ZI\,663/2-1).
IRS acknowledges support from the Royal Society and the Leverhulme Trust.

This work is based on observations with the Hale Telescope at Palomar
Observatory, which is owned and operated by California Institute of Technology.
This research has made use of the NASA/IPAC Extragalactic Database
(NED), which is operated by the Jet Propulsion Laboratory, California Institute
of Technology, under contract with the National Aeronautics and Space
Administration.



\appendix

\section{Properties of the \textit{HST} subsample}\label{hstprop}

\begin{figure*}
\centerline{%
\includegraphics[width=1.0\textwidth]{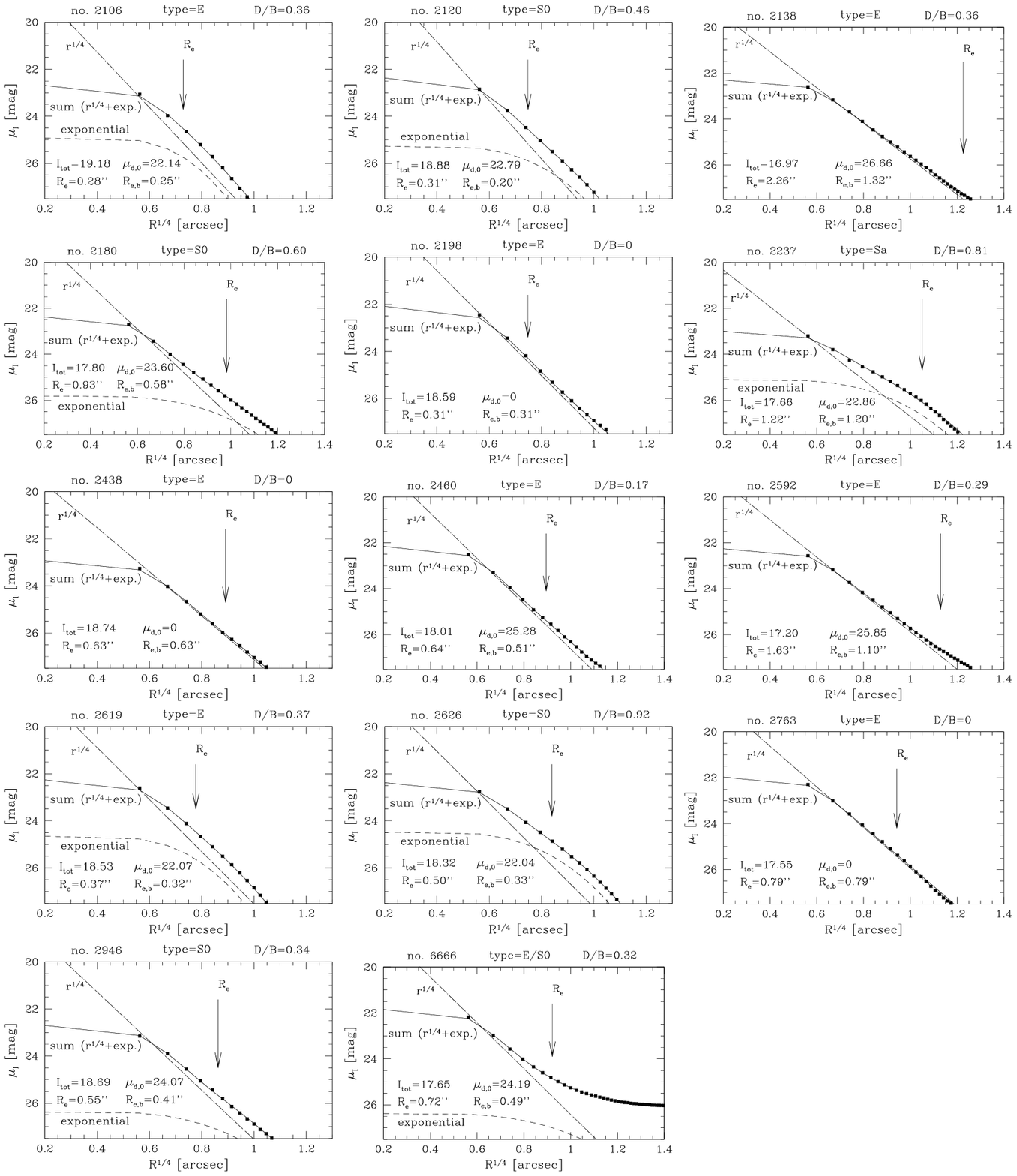}}
\caption{$I_{814}$ Surface brightness profiles
of galaxies residing within the \hst\ image with available spectral information.
The $I_{814}$ surface brightness magnitude (extinction A$_{F814W}$ corrected)
is plotted against the radius $R^{1/4}$ (in arcsec). Filled squares show
the observed profile, lines the different best models for the bulge
(de Vaucouleurs law), disk (exponential law) component fit and for a
combination of bulge and disk components (sum: $r^{1/4}+$exp.-law).
The arrow indicates the position of the effective radius $R_{e}$. Some
structural parameters are given (see text for details).}
\label{SBPHSTgals}
\end{figure*}

Surface brightness profiles of those galaxies situated within the \hst\ field of
Abell~2390 with available spectroscopic information are shown in
Fig.~\ref{SBPHSTgals}. For the surface brightnesses a correction for
galactic extinction was applied. In most cases the observed profile is best
fitted by a combination of a bulge (de Vaucouleurs law) and a disk
component (exponential law). Values for the total magnitude $I_{\mathrm{tot}}$,
the surface brightness of the disk $\mu_{\mathrm{d},0}$, the effective radius
$R_{e}$ and the effective radius of the bulge $R_{e,b}$, and the disk-to-bulge
ratio $D/B$ are listed. The arrow in each diagram indicates the position of
the effective radius $R_{e}$. Typical residuals $\Delta \mu_{I}$ between the
observed profile and the modelled fit are in the order of $\pm0.10$~mag at
$R^{1/4}\approx 1.3$ arcsec. Note that the disk fit indicated in
Fig.~\ref{SBPHSTgals} represents the disk component of the combined best fit
for the corresponding galaxies.

\begin{figure*}
\centerline{%
\includegraphics[width=0.80\textwidth]{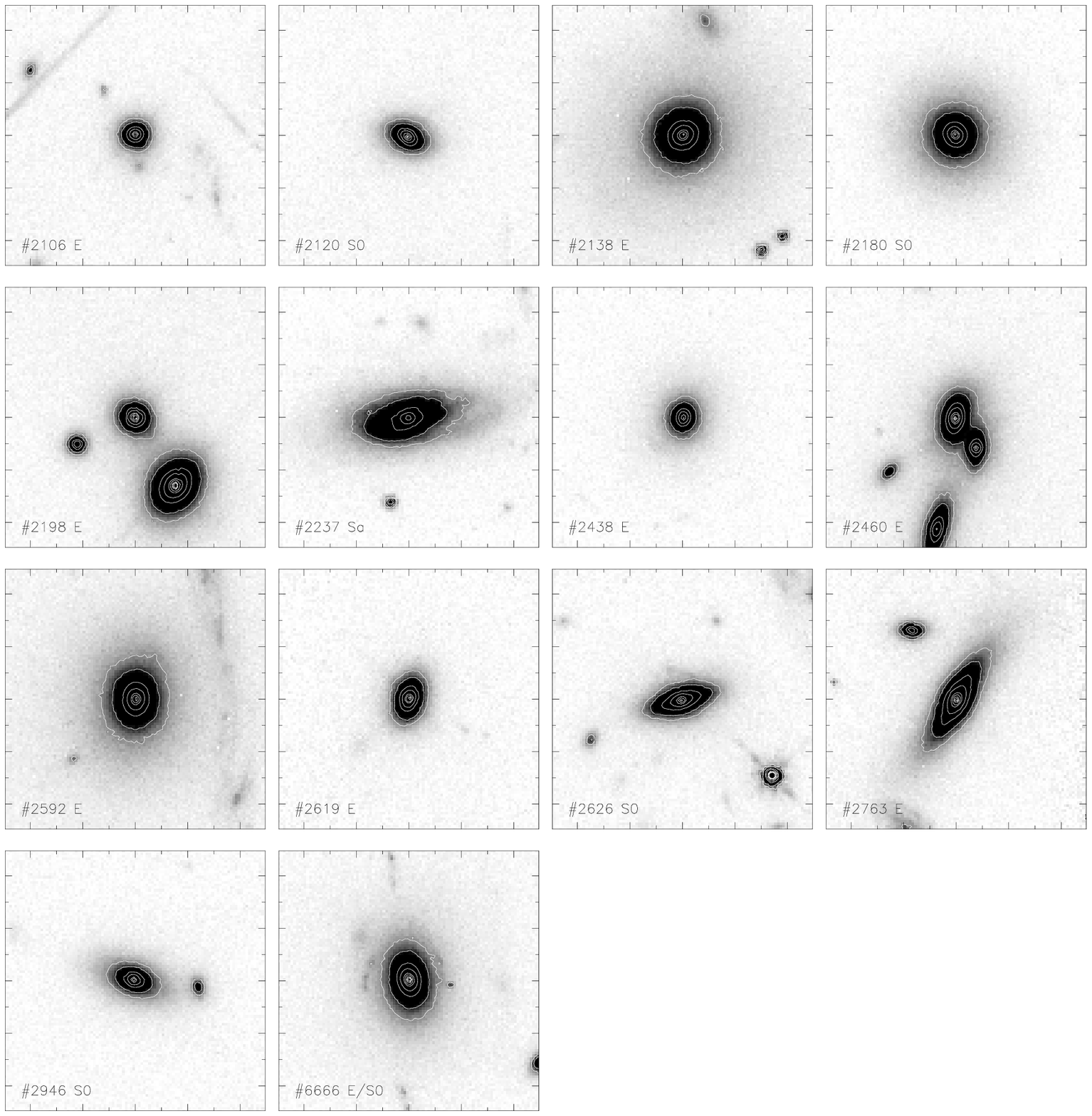}}
\caption{$10''\times10''$ HST/WFPC2 images of 14 galaxies with available
spectroscopic information which fall within the \textit{HST/WFPC2} field of
A\,2390. The labels give the galaxy ID and the visual morphology as listed in
Table~\ref{hstpar}. North is up and East to the left. The colour coding ranges
from white (sky-background at $\sim$26.30~mag arcsec$^{-2}$) to 
black ($\leq$23.11~mag arcsec$^{-2}$).
Isophotal contours are indicated between the range
$24.50\leq C_{i}\leq 21.94$~mag arcsec$^{-2}$, in increments of 0.5 magnitudes.}
\label{hstthumbs}
\end{figure*}

The single galaxy which could not be fitted (\#\,2933) does not reside close
to the edge of the WF4 chip but
is a late-type spiral galaxy (Sbc) with clear signs of spiral pattern.
Thumbnail images of the cluster members are enclosed in Fig.~\ref{hstthumbs}.
Objects were morphologically classified in two independent ways,
based on visual inspection as well as fitting routine output.
A visual classification was obtained by two of us, AF and IS separately,
resulting in the very similar morphologies.
The findings from the luminosity profile fitting provided a consistency check,
which resulted in the same classification scheme except for two objects
(\#\,2180: E (visual), S0 (fit) and \#\,2763: S0/Sa (visual), E (fit)).
Objects which are best described with an $r^{1/4}$-law are classified as an
elliptical (E); those galaxies for which an additional exponential component
yield in a slightly improved fit (without a dramatic change in structural
properties) are classified as E/S0; S0 galaxies are best approximated by a
combination of an $r^{1/4}$-law plus an exponential profile.

%

Remarks on special features on individual \hst\ objects of Abell~2390 are
listed in Table~\ref{hstrem}. The first column shows the galaxy ID, the second
indicates the morphology of the objects. In the third column, additional
information about special features regarding to the galaxy is given.

%
\begin{table*}
\centering
\begin{minipage}{100mm}
\caption{Noteworthy remarks on individual \hst\ objects in Abell~2390.}\label{hstrem}
\begin{tabular}{rcl}
\hline\hline
\noalign{\smallskip}
ID &  morp & comment \\
\noalign{\smallskip}
\hline
\noalign{\smallskip}
2120 & S0  &  signs of isophotal twists out to outer isophote \\
2138 & E   &  very extended low-surface brightness (LSB) environment \\
2198 & E   &  close to a bright neighbour \\
2237 & Sa  &  [O\,{\small III}]\,5007 and H$\beta$ emission, asymmetric LSB extension \\
2460 & E   &  lensed?, very close to S0 \\
2592 & E   &  next to giant arc \\
2763 & E   &  slight twisted low SB \\
2946 & S0  &  small companion \\ 
2933 & Sbc &  no member, no structural analysis, weird spiral structure, no interaction \\
\noalign{\smallskip}
\noalign{\hrule}
\end{tabular}
\end{minipage}
\end{table*}

\section{Photometric properties of cluster galaxies}\label{clusprop}

In this section we present the photometric parameters derived from the
ground-based $U$, $B$ and $I$-band Hale imaging of the cluster Abell~2390.
The galaxy ID can be found in the finding chart (Fig.~\ref{HALE}), 
$I_{\mathrm{tot}}$  is the total magnitude derived from the Hale images 
(SExtractor's {\sc best\_mag}), $U_{\mathrm{ap}}$, $B_{\mathrm{ap}}$, 
and $I_{\mathrm{ap}}$ are magnitudes within a circular aperture of
3.0\arcsec\ diameter measured in a seeing-matched $UBI$-image.
None of the given magnitudes are corrected for extinction.
$M_{r}$ denotes the absolute rest-frame Gunn $r$ magnitude.
The velocity dispersions of the galaxies $\sigma$ (in \kms) are not
aperture-corrected. Column $v$ denotes the heliocentric radial velocity
 (in \kms) and its error.
The galaxy with ID \#\,3824 is presumably an E$+$A candidate as its spectra
features strong H$\beta$ and H$\gamma$ Balmer lines. On the ground-based images
a disk is clearly visible but no further signs of spiral structure.
The object \#\,3038 is not included, because it is a foreground galaxy
at redshift of $z=0.1798$.

For the analysis we have also excluded the Sbc galaxy \#\,2933 and two other
objects (\#\,1507 and \#\,1639) because they all are background galaxies. Their
spectra suffer from low $S/N$ and since their G-band is affected by strong
sky-lines and the Mg feature falls into the telluric B-band no reliable
velocity dispersions could be derived. The properties of these galaxies and the
foreground galaxy \#\,3038 are shown in Table~\ref{nomem}.
During the observations of our target
galaxies four other galaxies fell into the slits by coincidence
(\#\,2222, \#\,5552, \#\,5553 and \#\,6666).
However, apart from the object \#\,6666, their spectra
are too faint ($S/N\la12$) to determine accurate velocity
dispersions. Thus, they were rejected from the final sample.
For the galaxy \#\,6666, we derived the absolute Gunn $r$-band magnitude 
using our \hst\ photometry and included it in our data set.

%
\begin{table*}\label{lastpage}
\centering
\begin{minipage}{175mm}
\caption{Properties of Abell~2390 cluster members.}\label{photprop}
\begin{tabular}{rcccccccccc}
\hline\hline
\noalign{\smallskip}
ID &  R.A. & Dec. & $v$ & $\sigma$ & $S/N$ & $I_{{\rm tot}}$ & $M_{r}$ &
$U_{\mathrm{ap}}$ & $B_{\mathrm{ap}}$ & $I_{\mathrm{ap}}$ \\
& \multicolumn{2}{c}{(J2000)} & [\kms] & [\kms] & & [mag] & [mag] & [mag]  &  [mag] &  [mag] \\
\noalign{\smallskip}
\hline
\noalign{\smallskip}
 803 & 21:53:37.851 & $+$17:38:26.50 & 67851.4$\pm 17$ & 148.0$\pm 15$ & 24 & 18.47 & -21.21 & 22.930$\pm 0.045$ & 22.346$\pm 0.012$ & 18.849$\pm 0.002$ \\
 869 & 21:53:46.482 & $+$17:38:19.79 & 67329.8$\pm 13$ & 106.9$\pm 15$ & 33 & 18.86 & -20.82 & 23.414$\pm 0.069$ & 22.687$\pm 0.014$ & 19.364$\pm 0.005$ \\
 943 & 21:53:40.247 & $+$17:37:59.77 & 69571.0$\pm 10$ & 204.6$\pm 08$ & 40 & 16.69 & -22.99 & 22.305$\pm 0.026$ & 21.588$\pm 0.008$ & 17.993$\pm 0.002$ \\
1254 & 21:53:49.782 & $+$17:39:14.32 & 68452.1$\pm 18$ & 106.4$\pm 18$ & 23 & 19.03 & -20.65 & 23.140$\pm 0.053$ & 22.714$\pm 0.015$ & 19.467$\pm 0.003$ \\
1345 & 21:53:47.789 & $+$17:39:25.65 & 70862.4$\pm 14$ & 129.1$\pm 12$ & 30 & 18.65 & -21.03 & 22.934$\pm 0.046$ & 22.456$\pm 0.014$ & 19.180$\pm 0.003$ \\
1493 & 21:53:46.838 & $+$17:39:49.11 & 68239.0$\pm 15$ & 154.4$\pm 16$ & 28 & 18.60 & -21.08 & 23.138$\pm 0.054$ & 22.490$\pm 0.013$ & 19.057$\pm 0.002$ \\
1712 & 21:53:54.375 & $+$17:40:06.60 & 68509.8$\pm 15$ & 095.4$\pm 17$ & 29 & 19.13 & -20.55 & 23.839$\pm 0.104$ & 23.184$\pm 0.025$ & 19.711$\pm 0.003$ \\
1787 & 21:53:36.848 & $+$17:40:34.89 & 68215.8$\pm 11$ & 153.3$\pm 14$ & 37 & 18.37 & -21.31 & 23.075$\pm 0.054$ & 22.296$\pm 0.012$ & 18.898$\pm 0.003$ \\
1843 & 21:53:48.694 & $+$17:40:24.96 & 69626.6$\pm 08$ & 213.5$\pm 07$ & 55 & 17.21 & -22.47 & 22.255$\pm 0.027$ & 21.485$\pm 0.008$ & 17.996$\pm 0.003$ \\
1893 & 21:53:39.944 & $+$17:40:37.38 & 72079.8$\pm 11$ & 119.4$\pm 13$ & 40 & 18.58 & -21.10 & 22.885$\pm 0.041$ & 22.229$\pm 0.010$ & 18.966$\pm 0.003$ \\
1941 & 21:53:45.708 & $+$17:40:43.92 & 68936.3$\pm 11$ & 166.6$\pm 14$ & 37 & 18.32 & -21.37 & 23.255$\pm 0.064$ & 22.394$\pm 0.014$ & 18.867$\pm 0.002$ \\
1959 & 21:53:43.856 & $+$17:40:49.44 & 69561.5$\pm 14$ & 159.7$\pm 16$ & 29 & 18.62 & -21.07 & 23.047$\pm 0.047$ & 22.385$\pm 0.013$ & 18.935$\pm 0.002$ \\
1977 & 21:53:40.606 & $+$17:40:50.81 & 69353.9$\pm 09$ & 212.9$\pm 11$ & 48 & 18.40 & -21.28 & 22.823$\pm 0.042$ & 22.249$\pm 0.011$ & 18.912$\pm 0.003$ \\
1983 & 21:53:48.952 & $+$17:40:46.37 & 68611.0$\pm 12$ & 138.8$\pm 14$ & 35 & 18.53 & -21.15 & 23.030$\pm 0.050$ & 22.352$\pm 0.012$ & 18.924$\pm 0.002$ \\
2054 & 21:53:56.603 & $+$17:40:38.70 & 67758.9$\pm 14$ & 246.6$\pm 10$ & 34 & 17.43 & -22.25 & 22.816$\pm 0.044$ & 22.019$\pm 0.010$ & 18.468$\pm 0.002$ \\
2057 & 21:53:40.518 & $+$17:41:03.88 & 70311.6$\pm 18$ & 108.8$\pm 18$ & 24 & 19.21 & -20.47 & 23.759$\pm 0.100$ & 22.947$\pm 0.019$ & 19.468$\pm 0.004$ \\
2106 & 21:53:37.624 & $+$17:41:09.49 & 70303.6$\pm 16$ & 131.0$\pm 16$ & 26 & 18.66 & -21.02 & 23.616$\pm 0.081$ & 22.933$\pm 0.020$ & 19.604$\pm 0.005$ \\
2120 & 21:53:39.159 & $+$17:41:08.84 & 68348.1$\pm 13$ & 127.2$\pm 15$ & 34 & 18.69 & -20.99 & 23.632$\pm 0.081$ & 22.840$\pm 0.021$ & 19.310$\pm 0.004$ \\
2126 & 21:53:46.558 & $+$17:40:56.81 & 68688.5$\pm 10$ & 172.7$\pm 10$ & 40 & 18.01 & -21.67 & 22.759$\pm 0.041$ & 22.074$\pm 0.011$ & 18.707$\pm 0.002$ \\
2138 & 21:53:36.158 & $+$17:41:12.92 & 73905.5$\pm 07$ & 150.0$\pm 08$ & 54 & 17.45 & -22.23 & 22.493$\pm 0.030$ & 21.737$\pm 0.008$ & 18.387$\pm 0.002$ \\
2161 & 21:53:50.345 & $+$17:40:58.65 & 65805.4$\pm 08$ & 153.9$\pm 09$ & 55 & 17.28 & -22.40 & 22.028$\pm 0.021$ & 21.373$\pm 0.006$ & 18.193$\pm 0.002$ \\
2169 & 21:53:51.994 & $+$17:40:58.94 & 69318.7$\pm 11$ & 185.1$\pm 10$ & 39 & 16.99 & -22.69 & 22.983$\pm 0.052$ & 22.288$\pm 0.013$ & 18.848$\pm 0.002$ \\
2180 & 21:53:33.586 & $+$17:41:27.44 & 68429.8$\pm 09$ & 146.2$\pm 10$ & 49 & 18.07 & -21.61 & 23.148$\pm 0.056$ & 22.262$\pm 0.011$ & 18.674$\pm 0.002$ \\
2195 & 21:53:43.144 & $+$17:41:19.67 & 71527.1$\pm 17$ & 128.0$\pm 18$ & 25 & 19.20 & -20.48 & 23.654$\pm 0.085$ & 22.899$\pm 0.017$ & 19.692$\pm 0.003$ \\
2198 & 21:53:37.641 & $+$17:41:24.93 & 70010.9$\pm 14$ & 135.0$\pm 15$ & 31 & 18.84 & -20.84 & 23.584$\pm 0.081$ & 22.589$\pm 0.014$ & 19.000$\pm 0.002$ \\
2237 & 21:53:31.360 & $+$17:41:34.36 & 73707.2$\pm 15$ & 147.1$\pm 12$ & 29 & 17.83 & -21.85 & 22.231$\pm 0.025$ & 21.693$\pm 0.007$ & 18.662$\pm 0.002$ \\
2373 & 21:53:45.730 & $+$17:41:32.35 & 68244.0$\pm 12$ & 156.3$\pm 11$ & 34 & 17.58 & -22.11 & 22.816$\pm 0.039$ & 21.913$\pm 0.009$ & 18.577$\pm 0.002$ \\
2438 & 21:53:31.878 & $+$17:41:59.61 & 71101.3$\pm 13$ & 108.6$\pm 15$ & 30 & 19.04 & -20.64 & 23.549$\pm 0.077$ & 22.787$\pm 0.016$ & 19.483$\pm 0.003$ \\
2453 & 21:53:21.110 & $+$17:42:06.57 & 69447.8$\pm 07$ & 137.3$\pm 08$ & 60 & 17.27 & -22.41 & 22.565$\pm 0.047$ & 21.845$\pm 0.009$ & 18.200$\pm 0.002$ \\
2460 & 21:53:38.369 & $+$17:41:47.78 & 69519.3$\pm 11$ & 234.5$\pm 10$ & 40 & 17.78 & -21.90 & 23.125$\pm 0.054$ & 22.194$\pm 0.011$ & 18.582$\pm 0.002$ \\
2511 & 21:53:28.182 & $+$17:42:12.63 & 69317.2$\pm 21$ & 139.8$\pm 18$ & 21 & 19.07 & -20.61 & 23.795$\pm 0.098$ & 22.965$\pm 0.019$ & 19.504$\pm 0.003$ \\
2537 & 21:53:42.673 & $+$17:41:53.94 & 67004.5$\pm 11$ & 188.8$\pm 13$ & 35 & 18.03 & -21.65 & 22.854$\pm 0.043$ & 22.182$\pm 0.011$ & 18.808$\pm 0.002$ \\
2592 & 21:53:34.514 & $+$17:41:57.74 & 68966.4$\pm 09$ & 174.1$\pm 11$ & 45 & 17.47 & -22.21 & 22.870$\pm 0.044$ & 21.919$\pm 0.009$ & 18.444$\pm 0.002$ \\
2619 & 21:53:35.931 & $+$17:42:13.22 & 69424.2$\pm 14$ & 207.9$\pm 14$ & 30 & 18.74 & -20.94 & 23.535$\pm 0.076$ & 22.606$\pm 0.014$ & 19.011$\pm 0.002$ \\
2626 & 21:53:34.492 & $+$17:42:14.28 & 67821.9$\pm 13$ & 177.4$\pm 12$ & 32 & 18.47 & -21.21 & 23.331$\pm 0.063$ & 22.371$\pm 0.012$ & 18.890$\pm 0.002$ \\
2763 & 21:53:31.400 & $+$17:42:28.91 & 65665.2$\pm 10$ & 272.5$\pm 12$ & 51 & 17.64 & -22.05 & 22.792$\pm 0.042$ & 21.889$\pm 0.009$ & 18.354$\pm 0.002$ \\
2946 & 21:53:35.048 & $+$17:42:49.44 & 72296.1$\pm 12$ & 119.6$\pm 15$ & 34 & 18.81 & -20.87 & 23.477$\pm 0.070$ & 22.834$\pm 0.016$ & 19.280$\pm 0.003$ \\
3028 & 21:53:25.715 & $+$17:43:40.70 & 70038.0$\pm 16$ & 143.1$\pm 11$ & 27 & 18.44 & -21.24 & 23.481$\pm 0.083$ & 22.607$\pm 0.014$ & 19.095$\pm 0.002$ \\
3053 & 21:53:36.433 & $+$17:44:13.33 & 68419.5$\pm 11$ & 227.0$\pm 13$ & 39 & 17.94 & -21.74 & 22.886$\pm 0.044$ & 22.004$\pm 0.009$ & 18.502$\pm 0.002$ \\
3060 & 21:53:30.898 & $+$17:44:45.62 & 69644.6$\pm 21$ & 147.3$\pm 16$ & 21 & 18.58 & -21.10 & 23.573$\pm 0.085$ & 22.724$\pm 0.017$ & 19.265$\pm 0.003$ \\
3201 & 21:53:22.506 & $+$17:44:10.77 & 67835.1$\pm 15$ & 175.6$\pm 17$ & 29 & 19.13 & -20.55 & 24.010$\pm 0.167$ & 23.171$\pm 0.022$ & 19.598$\pm 0.003$ \\
3473 & 21:53:25.146 & $+$17:44:16.95 & 66780.5$\pm 08$ & 140.4$\pm 10$ & 49 & 18.06 & -21.63 & 23.143$\pm 0.067$ & 22.236$\pm 0.014$ & 18.642$\pm 0.002$ \\
3529 & 21:53:30.135 & $+$17:43:57.75 & 71026.3$\pm 13$ & 125.7$\pm 10$ & 32 & 18.15 & -21.53 & 23.154$\pm 0.060$ & 22.336$\pm 0.012$ & 18.802$\pm 0.002$ \\
3760 & 21:53:47.760 & $+$17:42:45.52 & 72378.9$\pm 10$ & 147.7$\pm 12$ & 39 & 17.92 & -21.76 & 22.724$\pm 0.038$ & 21.968$\pm 0.009$ & 18.665$\pm 0.003$ \\
3805 & 21:53:27.091 & $+$17:43:36.42 & 69601.5$\pm 11$ & 227.7$\pm 07$ & 41 & 17.42 & -22.26 & 22.791$\pm 0.044$ & 21.870$\pm 0.009$ & 18.189$\pm 0.002$ \\
3814 & 21:53:28.668 & $+$17:42:52.37 & 68365.3$\pm 09$ & 220.2$\pm 10$ & 49 & 17.19 & -22.49 & 22.698$\pm 0.045$ & 21.872$\pm 0.009$ & 18.238$\pm 0.002$ \\
3824 & 21:53:27.658 & $+$17:45:06.43 & 66326.5$\pm 06$ & 147.4$\pm 08$ & 65 & 18.13 & -21.55 & 22.110$\pm 0.025$ & 21.679$\pm 0.007$ & 18.488$\pm 0.002$ \\
6666 & 21:53:39.059 & $+$17:42:59.13 & 67163.3$\pm 15$ & 191.3$\pm 13$ & 28 & 17.68 & -22.00 &      --         &	--	     &      --  	 \\
\noalign{\smallskip}
\noalign{\hrule}
\end{tabular}
\end{minipage}
\end{table*}
%

%
\begin{table*}
\centering
\begin{minipage}{175mm}
\caption{Properties of non cluster members.}\label{nomem}
\begin{tabular}{rcccccccccc}
\hline\hline
\noalign{\smallskip}
ID &  R.A. & Dec. & $v$ & $\sigma$ & $S/N$ & $I_{{\rm tot}}$ & $M_{r}$ &
$U_{\mathrm{ap}}$ & $B_{\mathrm{ap}}$ & $I_{\mathrm{ap}}$ \\
& \multicolumn{2}{c}{(J2000)} & [\kms] & [\kms] & & [mag] & [mag] & [mag]  &  [mag] &  [mag] \\
\noalign{\smallskip}
\hline
\noalign{\smallskip}
1507 & 21:53:45.512 & $+$17:39:57.40 & 98182.2$\pm 50$ &      --	 & -- & 19.52 & -20.16 & 23.778$\pm 0.091$ & 22.735$\pm 0.015$ & 19.645$\pm 0.003$ \\
1639 & 21:53:57.073 & $+$17:39:58.11 & 97402.8$\pm 50$ &      --	 & -- & 19.46 & -20.22 & 24.032$\pm 0.121$ & 22.976$\pm 0.020$ & 19.580$\pm 0.003$ \\
2933 & 21:53:32.404 & $+$17:42:48.76 & 119347.6$\pm 50$ &     --	 & -- & 19.47 & -20.21 & 23.688$\pm 0.092$ & 23.221$\pm 0.024$ & 19.979$\pm 0.004$ \\
3038 & 21:53:27.845 & $+$17:44:23.82 & 53904.0$\pm 08$ & 211.0$\pm 09$   & 50 & 17.77 & -21.91 & 22.197$\pm 0.026$ & 21.537$\pm 0.006$ & 18.249$\pm 0.002$ \\
\noalign{\smallskip}
\noalign{\hrule}
\end{tabular}
\end{minipage}
\end{table*}
%


\bsp


\begin{thebibliography}{83}


\bibitem[\protect\astroncite{Barger et al.}{1996}]{BAECSS96}
Barger, A.~J., Aragon-Salamanca, A., Ellis, R.~S., Couch, W.~J., Smail, I.,
Sharples, R.~M.
\newblock 1996, MNRAS, 279, 1

\bibitem[\protect\astroncite{Barnes \& Hernquist}{1992}]{BH92}
Barnes, J.~E., \& Hernquist, L. 1992,
\newblock ARA\&A 30, 705

\bibitem[\protect\astroncite{Baugh, Cole \& Frenk}{1996}]{BCF96}
Baugh, C.~M., Cole, S., Frenk, C.~S. 1996,
\newblock MNRAS, 283, 1361

\bibitem[\protect\citeauthoryear{Bender}{1988}]{Ben88}
Bender, R. 1988,
\newblock A\&A, 193, L7

\bibitem[\protect\astroncite{Bender et al.}{1988}]{BDM88}
Bender, R., D\"obereiner, S., \& M\"ollenhoff, C. 1988,
\newblock A\&AS, 74, 385

\bibitem[\protect\astroncite{Bender \& M{\"o}llenhoff}{1987}]{BM87}
Bender, R., M{\"o}llenhoff, C. 1987,
\newblock A\&A, 177, 71

\bibitem[\protect\citeauthoryear{Bender}{1990}]{Ben:90}
Bender, R., 1990,
\newblock A\&A, 229, 441

\bibitem[\protect\citeauthoryear{Bender, Burstein \& Faber}{1992}]{BBF92}
Bender, R., Burstein, D., Faber, S.~M. 1992,
\newblock ApJ, 399, 462

\bibitem[\protect\citeauthoryear{Bender, Burstein \& Faber}{1993}]{BBF93}
Bender, R., Burstein, D., Faber, S.~M. 1993,
\newblock ApJ, 411, 153

\bibitem[\protect\citeauthoryear{Bernardi et al.}{1998}]{Ber98}
Bernardi, M., Renzini, A., da Costa, L.~N., Wegner, G., Alonso, M.~V.,
Pellegrini, P.~S., Rit\'{e}, C., Willmer, C.~N.~A. 1998,
\newblock ApJ, 508, L143

\bibitem[\protect\citeauthoryear{Bernardi et al.}{2003}]{SDSSIIIFP03}
Bernardi, M. et al. 2003
\newblock AJ, 125, 1866

\bibitem[\protect\citeauthoryear{Bertin \& Arnouts}{1996}]{BA96}
Bertin, E.,  Arnouts, S., 1996,
\newblock A\&AS, 117, 393

\bibitem[\protect\citeauthoryear{Bruzual \& Charlot}{1993}]{BC93}
Bruzual, G.~A.,  Charlot, S. 1993,
\newblock ApJ, 405, 538

\bibitem[\protect\citeauthoryear{Caon et al.}{1993}]{Cao93}
Caon, N., Capaccioli, M., D'Onofrio, M. 1993,
\newblock MNRAS, 265, 1013

\bibitem[\protect\citeauthoryear{Carlberg et al.}{1996}]{Car:96}
Carlberg, R.~G., Yee, H.~K.~C., Ellingson, E., Abraham, R., Gravel, P., Morris,
S., Pritchet, C.~J. 1996,
\newblock ApJ, 462,~32

\bibitem[\protect\citeauthoryear{Cole et al.}{2000}]{Col:00}
Cole, S., Lacey, C.~G., Baugh, C.~M., Frenk, C.~S. 2000
\newblock MNRAS, 319, 168

\bibitem[\protect\citeauthoryear{de Carvalho \& Djorgovski}{1992}]{deCD:92}
de Carvalho, R.~R., Djorgovski, S. 1992,
\newblock ApJ, 389, L49

\bibitem[\protect\citeauthoryear{Djorgovski \& Davis}{1987}]{DD87}
Djorgovski, S.,  Davis, M. 1987,
\newblock ApJ, 313, 59

\bibitem[\protect\citeauthoryear{Dressler et al.}{1987}]{Dre:87}
Dressler, A., Lynden-Bell, D., Burstein, D., Davies, R.~L., Faber, S.~M.,
  Terlevich, R., Wegner, G. 1987,
\newblock ApJ, 313, 42

\bibitem[\protect\citeauthoryear{Dressler et al.}{1997}]{DOCSE97}
Dressler, A., Oemler Jr., A., Couch, W.~J., Smail, I., Ellis, R.~S.,
  Barger, A., Butcher, H., Poggianti, B.~M., Sharples, R.~M. 1997,
\newblock ApJ, 490, 577

\bibitem[\protect\citeauthoryear{Ellis et al.}{1997}]{E97}
Ellis, R.~S., Smail, I., Dressler, A., Couch, W.~J., Oemler, A.~J., Butcher,
H., Sharples, R.~M. 1997,
\newblock ApJ, 483, 582

\bibitem[\protect\citeauthoryear{Faber \& Jackson}{1976}]{FJ76}
Faber, S.~M., Jackson, R.~E. 1976,
\newblock ApJ, 204, 668

\bibitem[\protect\citeauthoryear{Faber et al.}{1987}]{Fab:87}
Faber, S.~M., Dressler, A., Davies, R.~L., Burstein, D., Lynden-Bell, D.,
Terlevich, R., Wegner, G. 1987, Nearly Normal Galaxies. 
\newblock Springer, New York, p. 175

\bibitem[\protect\citeauthoryear{Faber et al.}{1989}]{Fab:89}
Faber, S.~M., Wegner, G., Burstein, D., Davies, R.~L., Dressler,
A., Lynden-Bell, D., Terlevich, R.~J. 1989
\newblock ApJS, 69, 763

\bibitem[\protect\citeauthoryear{Fritz et al.}{2004}]{Fri:04}
Fritz, A., Ziegler, B.~L., Bower, R.~G., Smail, I., \& Davies, R.~L. 2004,
\newblock in \textit{``Clusters of Galaxies: Probes of Cosmological
   Structure and Galaxy Evolution,''} eds. J.~S. Mulchaey, A. Dressler,
   and A. Oemler. (Pasadena: Carnegie Observatories,
http://www.ociw.edu/ociw/symposia/series/\\symposium3/proceedings.html).

\bibitem[\protect\citeauthoryear{Fukugita et al.}{1995}]{FSI95}
Fukugita, M., Shimasaku, K.,  Ichikawa, T. 1995,
\newblock PASP, 107, 945

\bibitem[\protect\citeauthoryear{Girardi et al.}{1998}]{GGMMB98}
Girardi, M., Giuricin, G., Mardirossian, F., Mezzetti, M., \& Boschin, W. 1998,
\newblock ApJ, 505, 74

\bibitem[\protect\citeauthoryear{Graham et al.}{1996}]{Gra:96}
Graham, A., Lauer, T.~R., Colless, M., Postman, M. 1996,
\newblock ApJ, 465, 534

\bibitem[\protect\citeauthoryear{Graham \& Colless}{1997}]{GC97}
Graham, A., Colless, M. 1997,
\newblock  MNRAS, 287, 221

\bibitem[\protect\citeauthoryear{Guzm\'{a}n, Lucey \& Bower}{1993}]{Guz:93}
Guzm\'{a}n, R., Lucey, J.~R., \& Bower, R.~G. 1993,
\newblock MNRAS, 265, 731

\bibitem[\protect\citeauthoryear{Hill et al.}{1998}]{Hil:98}
Hill, R.~J. et al., 1998,
\newblock ApJ, 496, 648

\bibitem[\protect\citeauthoryear{Holtzman et al.}{1995}]{Hol:95}
Holtzman, J.~A., Burrows, C.~J., Casertano, S., Hester, J.~J.,
Trauger, J.~T., Watson, A.~M.,  Worthey, G. 1995,
\newblock PASP, 107, 1065

\bibitem[\protect\citeauthoryear{Horne}{1986}]{Horne86}
Horne, K. 1986,
\newblock PASP, 98, 609

\bibitem[\protect\citeauthoryear{James \& Mobasher}{1999}]{JM99}
James, P.~A., Mobasher, B. 1999,
\newblock MNRAS, 306, 199

\bibitem[\protect\citeauthoryear{J{\o}rgensen}{1999}]{Joerg99}
J{\o}rgensen, I. 1999,
\newblock MNRAS, 306, 607 (J99)

\bibitem[\protect\citeauthoryear{J{\o}rgensen et al.}{1999}]{JFHD99}
J{\o}rgensen, I., Franx, M., Hjorth, J., van Dokkum, P.~G. 1999,
\newblock MNRAS, 308, 833

\bibitem[\protect\citeauthoryear{J{\o}rgensen et al.}{1995}]{JFK95b}
J{\o}rgensen, I., Franx, M.,  Kj{\ae}rgaard, P. 1995,
\newblock MNRAS, 273, 1097 (J99)

\bibitem[\protect\citeauthoryear{J{\o}rgensen et al.}{1996}]{JFK96}
J{\o}rgensen, I., Franx, M.,  Kj{\ae}rgaard, P. 1996,
\newblock MNRAS, 280, 167

\bibitem[\protect\citeauthoryear{Kauffmann \& Charlot}{1998}]{KS:98}
Kauffmann, G., \& Charlot, S. 1998,
\newblock MNRAS, 294, 705

\bibitem[\protect\citeauthoryear{Kelson et al.}{2000a}]{KIDF00a}
Kelson, D.~D., Illingworth, G.~D., van Dokkum, P.~G.,  Franx, M.,
  2000a,
\newblock ApJ, 531, 137

\bibitem[\protect\citeauthoryear{Kelson et al.}{2000b}]{KIDF00}
Kelson, D.~D., Illingworth, G.~D., van Dokkum, P.~G.,  Franx, M.,
  2000b,
\newblock ApJ, 531, 184

\bibitem[\protect\citeauthoryear{Kelson et al.}{1997}]{KDFIF97}
Kelson, D.~D., van Dokkum, P.~G., Franx, M., Illingworth, G.~D.,
Fabricant, D. 1997,
\newblock ApJ, 478, L13

\bibitem[\protect\citeauthoryear{Kinney et al.}{1996}]{KCBMSS96}
Kinney, A.~L., Calzetti, D., Bohlin, R.~C., McQuade, K.,
Storchi-Bergmann, T., Schmitt, H.~R. 1996,
\newblock ApJ, 467, 38

\bibitem[\protect\citeauthoryear{Kodama et al.}{2001}]{KSNOB01}
Kodama, T., Smail, I., Nakata, F., Okamura, S., Bower, R.~G. 2001,
\newblock ApJ, 562, L9

\bibitem[\protect\citeauthoryear{Kormendy}{1977}]{Korme77}
Kormendy, J., 1977,
\newblock ApJ, 218, 333

\bibitem[\protect\citeauthoryear{Kormendy \& Bender}{1996}]{KB1996}
Kormendy, J., \& Bender, R. 1996,
\newblock ApJ, 464, L119

\bibitem[\protect\citeauthoryear{Kuntschner et al.}{2002}]{KSCD02}
Kuntschner, H., Smith, R.~J., Colless, M., Davies, R.~L., Kaldare, R.,
Vazdekis, A. 2002
\newblock MNRAS, 337, 172

\bibitem[\protect\citeauthoryear{La Barbera et al.}{2002}]{LBBMMC02}
La Barbera, F., Busarello, G., Merluzzi, P., Massarotti, M.,
Capaccioli, M. 2002,
\newblock ApJ, 571, 790

\bibitem[\protect\citeauthoryear{Larson}{1975}]{Lar:75}
Larson, R.~B. 1975,
\newblock MNRAS, 173, 671

\bibitem[\protect\citeauthoryear{Le Borgne et al.}{1991}]{LeB:91}
Le Borgne, J.~F., Mathez, G., Mellier, Y., Pello, R., Sanahuja, B. \& Soucail,
G. 1991,
\newblock A\&AS, 88, 133

\bibitem[\protect\citeauthoryear{Le Borgne et al.}{1992}]{LPS92}
Le Borgne, J.~F., Pello, R.,  Sanahuja, B. 1992,
\newblock A\&AS, 95, 87

\bibitem[\protect\citeauthoryear{Mehlert et al.}{2003}]{MTS03}
Mehlert, D., Thomas, D., Saglia, R.~P., Bender, R., Wegner, G. 2003,
\newblock A\&A, 407, 423

\bibitem[\protect\citeauthoryear{Menanteau et al.}{2001}]{MAE:01}
Menanteau, F., Abraham, R.~G., Ellis, R.~S. 2001,
\newblock MNRAS, 322, 1

\bibitem[\protect\astroncite{Moore et~al.}{1996}]{MLKDO96}
Moore, B., Katz, N., Lake, G., Dressler, A., \& Oemler, A. 1996,
\newblock Nature 379, 613

\bibitem[\protect\astroncite{M{\"o}ller et~al.}{2001}]{Moe01}
M{\"o}ller, C.~S., Fritze-v.Alvensleben, U., Fricke, K.~J.,
\& Calzetti, D. 2001,
\newblock Ap\&SS, 276, 799

\bibitem[\protect\astroncite{Naab et al.}{1999}]{NB99}
Naab, T., Burkert, A., Hernquist, L. 1999,
\newblock ApJ, 523, L133

\bibitem[\protect\astroncite{Naab \& Burkert}{2003}]{NB03}
Naab, T., Burkert, A. 2003,
\newblock ApJ, 597, 893

\bibitem[\protect\citeauthoryear{Pahre et al.}{1998a}]{PDC98a}
Pahre, M.~A., Djorgovski, S.~G., \& de Carvalho, R.~R. 1998,
\newblock AJ, 116, 1591

\bibitem[\protect\citeauthoryear{Poggianti et al.}{1999}]{PSDCB99}
Poggianti, B.~M., Smail, I., Dressler, A., Couch, W.~J., Barger, A.,
  Butcher, H., Ellis, R.~S.,  Oemler~Jr., A. 1999,
\newblock ApJ, 518, 576

\bibitem[\protect\citeauthoryear{Rusin et al.}{2003}]{RKFKM03}
Rusin, D. et al. 2003,
\newblock  ApJ, 587, 143

\bibitem[\protect\citeauthoryear{Saglia et al.}{1993}]{SBD93}
Saglia, R.~P., Bender, R., Dressler, A. 1993,
\newblock A\&A, 279, 75

\bibitem[\protect\citeauthoryear{Saglia et al.}{1997a}]{Sag:97a}
Saglia, R.~P., Bertschinger, E., Baggley, G., Burstein, D., Colless,
  M., Davies, R.~L., McMahan Jr., R.~K.,  Wegner, G., 1997a,
\newblock ApJS, 109, 79

\bibitem[\protect\citeauthoryear{Saglia et al.}{1997b}]{Sag:97b}
Saglia, R.~P., Burstein, D., Baggley, G., Davies, R.~L., Bertschinger, E.,
  Colless, M., McMahan Jr., R.~K.,  Wegner, G., 1997b,
\newblock MNRAS, 292, 499

\bibitem[\protect\citeauthoryear{S\'{a}nchez-Bl\'{a}zquez et al.}{2003}]{SBGCCG03}
S\'{a}nchez-Bl\'{a}zquez, P., Gorgas, J., Cardiel, N., Cenarro, J.,
Gonz\'{a}lez, J.~J. 2003,
\newblock ApJ, 590, L91

\bibitem[\protect\citeauthoryear{Schade et al.}{1999}]{SLCE99}
Schade, D. et al. 1999,
\newblock ApJ, 525, 31

\bibitem[\protect\citeauthoryear{Schlegel et al.}{1998}]{SFD98}
Schlegel, D.~J., Finkbeiner, D.~P.,  Davis, M. 1998,
\newblock ApJ, 500, 525

\bibitem[\protect\citeauthoryear{Simard et al.}{2002}]{SWVSP02}
Simard, L. et al. 2002,
\newblock ApJS, 142, 1

\bibitem[\protect\citeauthoryear{Smail et al.}{1998}]{SEEB98}
Smail, I., Edge, A.~C., Ellis, R.~S., Blandford, R.~D. 1998,
\newblock MNRAS, 293, 124

\bibitem[\protect\citeauthoryear{Smail et al.}{2001}]{SKKSPFH01}
Smail, I., Kuntschner, H., Kodama, T., Smith, G.~P., Packham, C.,
Fruchter, A.~S., Hook, R.~N. 2001,
\newblock MNRAS, 323, 839

\bibitem[\protect\citeauthoryear{Stanford et al.}{1998}]{Sta:98}
Stanford, S.~A., Eisenhardt, P.~R., Dickinson,M.~E. 1998,
\newblock ApJ, 492, 461

\bibitem[\protect\citeauthoryear{Thuan \& Gunn}{1976}]{TG76}
Thuan, T.~X., Gunn, J.~E. 1976
\newblock PASP, 88, 543

\bibitem[\protect\citeauthoryear{Toomre}{1977}]{Too:77}
Toomre, A. 1977,
\newblock in {\it ``The Evolution of Galaxies and Stellar populations''}\/,
eds. Tinsley B.~T., Larson R.~B., Yale University Press, New Haven, p. 401

\bibitem[\protect\citeauthoryear{Treu et al.}{2001a}]{Treu01a}
Treu, T., Stiavelli, M., M{\o}ller, P., Casertano, S., \& Bertin, G. 2001a
\newblock MNRAS, 326, 221

\bibitem[\protect\citeauthoryear{Treu et al.}{2001b}]{Treu01b}
Treu, T., Stiavelli, M., Bertin, G., Casertano, S., \& M{\o}ller, P. 2001b
\newblock MNRAS, 326, 237

\bibitem[\protect\citeauthoryear{Treu et al.}{2002}]{TSCMB02}
Treu, T., Stiavelli, M., Casertano, S., M{\o}ller, P., Bertin, G. 2002,
\newblock ApJ, 564, L13

\bibitem[\protect\citeauthoryear{Treu et al.}{2003}]{Treu03}
Treu, T., Ellis, R.~S., Kneib, J.-P., Dressler, A., Smail,
I., Czoske, O., Oemler, A., Natarajan, P. 2003,
\newblock ApJ, 591, 53

\bibitem[\protect\citeauthoryear{Tran et al.}{2003}]{TSIF03}
Tran, K.~H., Simard, L., Illingworth, G., Franx, M. 2003,
\newblock ApJ, 590, 238

\bibitem[\protect\citeauthoryear{van Dokkum \& Franx}{1996}]{DF96}
van Dokkum, P.~G.,  Franx, M. 1996,
\newblock MNRAS, 281, 985

\bibitem[\protect\citeauthoryear{van Dokkum et al.}{1998}]{DFKI98}
van Dokkum, P.~G., Franx, M., Kelson, D.~D., Illingworth, G.~D. 1998,
\newblock ApJ, 504, L17

\bibitem[\protect\citeauthoryear{van Dokkum et al.}{2000}]{vD00}
van Dokkum, P.~G., Franx, M., Fabricant, D., Illingworth, G.~D.,
Kelson, D.~D. 2000,
\newblock ApJ, 541, 95

\bibitem[\protect\citeauthoryear{van Dokkum et al.}{2001}]{vD01}
van Dokkum, P.~G., Franx, M., Kelson, D.~D. Illingworth, G.~D. 2001,
\newblock ApJ, 553, L39

\bibitem[\protect\citeauthoryear{van Dokkum \& Franx}{2001}]{vD01b}
van Dokkum, P.~G., \& Franx, M. 2001,
\newblock ApJ, 553, 90

\bibitem[\protect\citeauthoryear{Wuyts et al.}{2004}]{WvDKFI04}
Wuyts, S., van Dokkum, P.~G., Kelson, D.~D., Franx, M., Illingworth, G.~D.
\newblock 2004, ApJ, 605, 677

\bibitem[\protect\citeauthoryear{Yang et al.}{2004}]{YZZea04}
Yang, Y., Zabludoff, A.~I., Zaritsky, D., Lauer, T.~R., Mihos, J.~C.
\newblock 2004, ApJ, 607, 258

\bibitem[\protect\citeauthoryear{Yee et al.}{1996}]{CNOCII}
Yee, H.~K.~C., Ellingson, E., Abraham, R.~G., Gravel, P., Carlberg, R.~G.,
Smecker-Hane, T.~A., Schade, D., Rigler, M. 1996,
\newblock ApJS, 102, 289

\bibitem[\protect\citeauthoryear{Ziegler \& Bender}{1997}]{ZB97}
Ziegler, B.~L., Bender, R. 1997,
\newblock MNRAS, 291, 527

\bibitem[\protect\citeauthoryear{Ziegler et al.}{1999}]{ZSBBGS99}
Ziegler, B.~L., Saglia, R.~P., Bender, R., Belloni, P., Greggio, L., Seitz, S.
  1999,
\newblock A\&A, 346, 13

\bibitem[\protect\citeauthoryear{Ziegler et al.}{2001}]{Z2001}
Ziegler, B.~L., Bower, R.~G., Smail, I., Davies, R.~L., Lee, D., 2001,
\newblock MNRAS, 325, 1571 (Z01)

 
\end{thebibliography}
\end{document}